\documentclass[11pt,a4paper]{article}

\pdfoutput=0
\usepackage{cite}
\usepackage{graphicx}
\usepackage{jheppub}
\usepackage{amsmath}
\usepackage{amssymb}
\usepackage{enumerate}
\usepackage{multirow}
\usepackage{lscape}
\usepackage{slashed}
\usepackage{axodraw4j}
\usepackage{pstricks}
\usepackage{color}

\graphicspath{ {figs/} }

\newcommand{\Graph}[2]{\vcenter{\hbox{\includegraphics[scale=#1]{#2}}}}

\let\OLDthebibliography\thebibliography
\renewcommand\thebibliography[1]{
  \OLDthebibliography{#1}
  \setlength{\parskip}{0pt}
  \setlength{\itemsep}{0pt plus 0.3ex}
}

\newenvironment{changemargin}[2]{%
  \begin{list}{}{%
    \setlength{\topsep}{0pt}%
    \setlength{\leftmargin}{#1}%
    \setlength{\rightmargin}{#2}%
    \setlength{\listparindent}{\parindent}%
    \setlength{\itemindent}{\parindent}%
    \setlength{\parsep}{\parskip}%
  }%
  \item[]}{\end{list}}
  
\def\beq{\begin{equation}}
\def\eeq{\end{equation}}
\def\bsp#1\esp{\begin{split}#1\end{split}}
\def\beqa{\begin{eqnarray}}
\def\eeqa{\end{eqnarray}}

\newcommand{\ep}{\epsilon}

\title{Baikov-Lee Representations Of Cut Feynman Integrals}

\author{Mark Harley,}
\author{Francesco Moriello,}
\author{Robert M. Schabinger}

\affiliation{Hamilton Mathematics Institute, School of Mathematics, Trinity College Dublin,\\ College Green, Dublin 2, Ireland}

\emailAdd{mharley.physics@gmail.com}
\emailAdd{moriello@maths.tcd.ie}
\emailAdd{schabr@maths.tcd.ie}

\abstract{We develop a general framework for the evaluation of $d$-dimensional cut Feynman integrals based on the Baikov-Lee representation of purely-virtual Feynman integrals. We implement the generalized Cutkosky cutting rule using Cauchy's residue theorem and identify a set of constraints which determine the integration domain. The method applies equally well to Feynman integrals with a unitarity cut in a single kinematic channel and to maximally-cut Feynman integrals. Our cut Baikov-Lee representation reproduces the expected relation between cuts and discontinuities in a given kinematic channel and furthermore makes the dependence on the kinematic variables manifest from the beginning. By combining the Baikov-Lee representation of maximally-cut Feynman integrals and the properties of periods of algebraic curves, we are able to obtain complete solution sets for the homogeneous differential equations satisfied by Feynman integrals which go beyond multiple polylogarithms. We apply our formalism to the direct evaluation of a number of interesting cut Feynman integrals.}

\begin{document}

\begin{flushright}
TCDMATH-17-12
\vspace*{-25pt}
\end{flushright}
\maketitle
\allowdisplaybreaks


\section{Introduction}
\label{sec:intro}
As has long been known, the evaluation of multi-loop Feynman integrals is an important component of many high-precision collider physics calculations. Even at the two-loop level, both purely-virtual and cut Feynman integrals often provide a remarkable computational challenge. This is particularly true if one proceeds analytically, and, as a consequence, a number of specialized techniques have been developed to aid in the evaluation of such integrals.  Most direct analytic integration techniques have traditionally employed some variant of the well-known Feynman (or Schwinger) parametric representation \cite{Feynman:1949zx} as a starting point in the purely-virtual case, due to the fact that it makes the dependence of the integral on Lorentz-invariant quantities manifest. The aim of this paper is to supply an analogous framework for cuts which is suitable not only for the direct calculation of real radiative master integrals, but also for the maximally-cut integrals relevant to the study of purely-virtual Feynman integrals in the method of differential equations~\cite{Kotikov:1990kg,Kotikov:1991hm,Kotikov:1991pm,Bern:1992em,Bern:1993kr,Remiddi:1997ny,Gehrmann:1999as}. As we shall see, our work is a natural out-growth of earlier work on the subject of integral reduction which we apply in a novel way to evaluate various types of cut Feynman integrals.

It is unfortunately the case that phenomenologically-relevant multi-loop integration by parts reductions \cite{Tkachov:1981wb,Chetyrkin:1981qh} often require dedicated effort and substantial resources to compute. Although it is probably fair to say that most recent higher-order perturbative calculations rely on some variant of Laporta's algorithm \cite{Laporta:2001dd,vonManteuffel:2012np,vonManteuffel:2014ixa,Smirnov:2014hma}, several other interesting and useful algorithms have been worked out and implemented over the years (see {\it e.g.} \cite{Gorishnii:1989gt,Baikov:1996iu,Gluza:2010ws,Lee:2012cn,Ruijl:2015aca,Ita:2015tya,Larsen:2015ped,Ueda:2016sxw,Abreu:2017idw,Abreu:2017xsl}). However, even very different approaches, such as the one advocated by Baikov \cite{Baikov:1996iu}, sometimes turn out to have ramifications for Laporta's method as well. Some time ago, the derivation of the Baikov formula for purely-virtual Feynman integrals was clarified by Lee and then used to great effect as a generating function for integration by parts relations~\cite{Lee:2010wea}. The product of his analysis, what we shall hereafter refer to as the Baikov-Lee representation, will be of great interest to us in this work. Larsen and Zhang argued in \cite{Larsen:2015ped} that evaluating the integration by parts relations generated by the Baikov representation on the support of various cuts dramatically improves the approach to integral reduction originally advocated by Gluza, Kajda, and Kosower \cite{Gluza:2010ws}. In fact, from their work and a similar study by Ita \cite{Ita:2015tya}, one can readily guess that the Baikov-Lee representation ought to offer a useful starting point for the evaluation of cut Feynman integrals.

Even though cut Feynman integrals play every bit as important a role as purely-virtual Feynman integrals in the perturbative computation of collider observables, methods for their direct evaluation have not been as thoroughly developed.\footnote{Certainly, we do not wish to suggest that such techniques do not exist. For instance, reference \cite{Anastasiou:2013srw} describes some pertinent traditional and state-of-the-art direct integration techniques in great detail. In fact, a setup which bears at least some rudimentary resemblance to the one discussed in this paper was developed for tree-level cross section calculations in $d = 4$ long ago~\cite{Kumar:1970cr}.} It is entirely possible that, at least in part, this state of affairs has persisted due to the fact that it is not completely trivial to write down a Baikov representation for Feynman integrals directly in Minkowski space. The issue is that the standard derivation of the Baikov-Lee representation (see {\it e.g.} \cite{Grozin:2011mt} for a detailed exposition) relies heavily upon Euclidean geometric intuition which does not immediately generalize. In this paper, we write down a simple recipe for the analytical continuation of the Euclidean Baikov-Lee formula by drawing an analogy to the more familiar situation that one encounters in the derivation of the Feynman parametric representation. To pass from uncut to cut propagators, we use sequential applications of Cauchy's residue theorem to implement a natural generalization of Cutkosky's cutting rule \cite{Cutkosky:1960sp,Lee:2012te}.\footnote{Due to the fact that the generalized Cutkosky rule of reference \cite{Lee:2012te} is actually written in terms of delta distributions and their derivatives, certain subtleties apply. To be consistent, one must first integrate out a complete set of scalar products localized by the distributions before attempting any of the more non-trivial integrations over scalar products. We would like to thank Ruth Britto for emphasizing this point to us.} Finally, the integration region is determined by analyzing the analytical structure of the integrand and applying the available constraints from physics. At each step of the calculation, one integrates between branch points of the integrand with respect to the current variable of integration. This procedure allows one to write down Baikov-Lee representations for a wide class of cut Feynman integrals. Crucially, this approach makes the dependence of the cut integral on the kinematic invariants of the problem manifest and eliminates the need to set up a convenient reference frame to carry out integrations over the components of the cut loop momenta.\footnote{To appreciate this point, we invite the reader to compare and contrast the maximally-cut Feynman integral calculation which appears in both references~\cite{Primo:2016ebd} and~\cite{Frellesvig:2017aai}.}

As mentioned above, the Baikov-Lee representation for cut Feynman integrals also has important applications to the more indirect method of differential equations. Writing differential equations with respect to the available kinematic parameters allows for the complete determination of a large class of Feynman integrals in terms of an appropriate set of iterated integrals, order-by-order in the parameter of dimensional regularization, $\ep$. For simple families of Feynman integrals with relatively few ratios of scales, the function space associated to Feynman integrals is spanned by iterated integrals with rational integrating factors. This set of iterated integrals, comprised of the well-known multiple polylogarithms \cite{LDanilevsky}, has been studied and popularized by many authors (\cite{Goncharov:1998kja,Remiddi:1999ew,Gehrmann:2000zt} to name a few). In the polylogarithmic case, it is always possible, by suitably choosing the basis of Feynman integrals~\cite{Kotikov:2010gf,Henn:2013pwa,Henn:2013tua}, to reduce the problem to one which admits an elementary formal solution in terms of Chen iterated integrals~\cite{Chen:1977oja}. Using the symbol-coproduct calculus~\cite{Goncharov:2005sla,Brown:2011ik,Duhr:2011zq,Duhr:2012fh}, performance-optimized solutions which can be readily interfaced with a Monte Carlo integration program may be constructed for the purposes of phenomenology (see {\it e.g.} \cite{Gehrmann:2015ora,vonManteuffel:2015msa,Bonciani:2015eua,vonManteuffel:2017myy}). Even through to weight four, this is not always straightforward to do in practice, despite the fact that the weight four function space has been studied extensively and is in principle well-understood~\cite{Frellesvig:2016ske}.

For general Feynman integrals, far more complicated analytic structures may appear; even simple-looking two-loop integrals which depend on sufficiently many kinematic variables may already involve elliptic polylogarithms and related functions~\cite{BrownLevin,Bloch:2013tra,Bloch:2014qca,Adams:2014vja,Adams:2015gva,Adams:2016xah}. In this context, maximally-cut Feynman integrals play an important role because they satisfy the homogeneous differential equations for the associated uncut Feynman integrals~\cite{Lee:2012te,Primo:2016ebd}.\footnote{Although, we will leave a detailed discussion for future work, there is nothing stopping us from applying this technique also to the differential equations satisfied by multi-scale cut Feynman integrals in the reverse-unitarity method~\cite{Anastasiou:2002yz,Anastasiou:2003yy,Anastasiou:2003ds}.} This property is particularly useful when Feynman integrals cannot be expressed in terms of multiple polylogarithms, and iterated integrals over special functions need to be considered~\cite{Caffo:1998du,Laporta:2004rb,Adams:2015ydq,Bloch:2016izu,Remiddi:2016gno,Bonciani:2016qxi,vonManteuffel:2017hms}. In fact, higher-order differential equations appear in non-polylogarithmic cases, and, at the present time, no general solution algorithm is known. Nevertheless, it was observed by Primo and Tancredi in~\cite{Primo:2016ebd} that, upon setting $\ep$ to zero, maximally-cut Feynman integrals can often be computed in closed form, allowing one to find at least a single homogeneous solution to the higher-order differential equations under consideration by direct integration. Provided that a complete, linearly independent set of homogeneous solutions can be found, the full solution can finally be determined using the variation of parameters technique. In our opinion, it is of great importance to supply an algorithm which comes up with not just one, but rather a complete set of homogeneous solutions. We provide a general prescription\footnote{A few days prior to the appearance of this paper, we became aware of a recent preprint, \cite{Primo:2017ipr}, which discusses many of the same technical issues for non-polylogarithmic Feynman integrals. In fact, the maximally-cut case was also discussed recently in the Baikov approach by yet another group \cite{Bosma:2017ens}, with, however, a different set of physical problems in mind.} and show that it allows for the straightforward construction of complete sets of homogeneous solutions for the non-polylogarithmic examples of Section \ref{sec:maxcut}.

The plan of this paper is as follows. In Section \ref{sec:genform}, we explain in detail how to write down Baikov-Lee representations for Feynman integrals cut in a single kinematic channel and how to work out an explicit description of the relevant integration domain. Our focus will be on the physical case relevant to cross section calculations because the procedure for maximally-cut Feynman integrals is closely analogous and has already been discussed in reference \cite{Frellesvig:2017aai}. Although we do not have a computer program which finds the integration limits for arbitrarily complicated cut Feynman integrals, we have a solid conceptual understanding which could in the future lead us to an explicit algorithm. In Section \ref{sec:stdcuts}, we go through a number of well-studied one- and two-loop examples of Feynman integrals cut in a single kinematic channel in order to give the reader a feeling for how explicit computations typically proceed when one adopts a Baikov-Lee representation as the starting point. To the best of our knowledge, our treatment of classical cut Feynman integrals in the Baikov-Lee representation is new and effectively extends the work of Frellesvig and  Papadopoulos~\cite{Frellesvig:2017aai} beyond the maximally-cut case. Although we work with generic values of the spacetime dimension for pedagogical purposes, our calculations strongly suggest that, in practice, an expansion in $\ep$ under the integral sign must be the way to go for all but the very simplest of cut Feynman integrals. 

In Section \ref{sec:maxcut}, we move on to maximally-cut Feynman integrals which evaluate to complete elliptic integrals. We discuss a general solution strategy applicable to many problems of practical interest and then demonstrate the general procedure by focusing on examples which are suitable for exposition. We emphasize in particular the utility of integrating out one loop at a time, as this generically leads to simpler Baikov-Lee representations. Finally, we conclude in Section \ref{sec:conc} and outline our plans for future research. We also include a number of appendices for pedagogical purposes and cross-checks. To streamline the exposition in the body of the paper, we summarize a number of purely mathematical results from the theory of hypergeometric-like functions in Appendix \ref{app:hypergeo}. In Appendix \ref{app:virtual}, we reproduce physical-region results from the literature for the uncut versions of the integrals considered in Section \ref{sec:stdcuts}. This allows the reader to easily verify our results using the classical relation between discontinuities and cuts in a given kinematic channel (see {\it e.g.} \cite{tHooft:1973wag,Gehrmann-DeRidder:2003pne,Abreu:2014cla}). Finally, in Appendix \ref{app:Baikov}, we evaluate a simple uncut Feynman integral using the Baikov-Lee setup to help the less familiar reader understand the relation between our prescriptions for cut Feynman integrals and the usual prescriptions for Baikov's method in the purely-virtual case.


\section{General formalism}
\label{sec:genform}
In this section, we define our notation, recall some results from the literature, and explain how we generalize the Baikov-Lee representation to the case of cut Feynman integrals. 
\subsection{Preliminaries}
Let us begin by discussing our notation for purely-virtual, $L$-loop Feynman integrals and recalling some useful facts about them. For the direct integration of purely-virtual Feynman integrals, a very common starting point is the Feynman (or Schwinger) parametric representation~(see {\it e.g.}~\cite{Smirnov:2004ym} for a detailed exposition). In many cases, it is convenient to write down the Feynman parametric representation in Euclidean space, treating all $n$ external momenta, $\{p_i\}$, on an equal footing by taking them all to be outgoing. In the most general case~\cite{Smirnov:2010hn}, it suffices to consider Feynman integrals of the form
\begin{align}
\label{eq:gen-virt-int-E}
  I_{\rm E} = \int\!{\rm d}^d k_1 \cdots \int\!{\rm d}^d k_L ~\prod_{\ell = 1}^N \left(Q_\ell^2(k_i,p_j) + m_\ell^2\right)^{-\nu_\ell},
\end{align}
where $Q_\ell(k_i,p_j)$ denotes the momentum of the $\ell$-th propagator and the $N$ propagators in (\ref{eq:gen-virt-int-E}) are linearly independent. It is often the case that one can profitably work with the Feynman representation for the all-plus metric and ultimately obtain results which differ from the Minkowski space results in an appropriate Euclidean kinematic region only by trivial phases.\footnote{Of course, certain assumptions must be satisfied. For a more in-depth discussion, see reference~\cite{Panzer:2015ida}.}

This approach has been used to great effect in recent years by Brown, Panzer, and others~\cite{Brown:2008um,Brown:2009ta,Panzer:2013cha,Panzer:2014gra,Panzer:2014caa,vonManteuffel:2014qoa,Panzer:2015ida,vonManteuffel:2015gxa}, culminating recently in an impressive calculation of the six-loop $\beta$ function in $\phi^4$ theory~\cite{Kompaniets:2016hct}. Working through the details of the straightforward derivation (see {\it e.g.} \cite{Smirnov:2004ym}), one finds the all-plus Feynman parametrization
\begin{align}
\label{eq:gen-parametric-rep-E}
  I_{\rm E} = \frac{\pi^{\frac{L d}{2}} \Gamma\big(\nu - \frac{L d}{2}\big)}{\prod_{i=1}^N \Gamma(\nu_i)} \Bigg[ \prod_{j=1}^N \int_0^{\infty} {\rm d} x_j \Bigg] \delta(1-x_N)\,\mathcal{U}_{\rm E}^{\nu - (L+1)d/2} \mathcal{F}_{\rm E}^{L d/2-\nu} \prod_{k=1}^N x_k^{\nu_k-1},
\end{align}
where $\mathcal{U}_{\rm E}$ and $\mathcal{F}_{\rm E}$ are respectively the first and second Symanzik polynomials~\cite{Bogner:2010kv} in Euclidean space and $\nu = \sum_{i=1}^N \nu_i$. However, for the evaluation of Feynman integrals relevant to the computation of collider observables, it is arguably more natural to work in Minkowski space from the very beginning, considering Feynman integrals of the form
\begin{align}
\label{eq:gen-virt-int-M}
  I_{\rm M} = \int\!{\rm d}^d k_1 \cdots \int\!{\rm d}^d k_L ~\prod_{\ell = 1}^N \left(Q_\ell^2(k_i,p_j) - m_\ell^2 + i 0\right)^{-\nu_\ell}
\end{align}
with a momentum flow suitable for the description of a scattering experiment.

As is well-known, the derivation of Eq. (\ref{eq:gen-parametric-rep-E}) goes through with minor modifications if one works directly in Minkowski space. The mostly-minus Feynman parameter representation has the form
\begin{changemargin}{-.175 cm}{0 cm}
\begin{align}
\label{eq:gen-parametric-rep-M}
  I_{\rm M} = \frac{i^L \pi^{\frac{L d}{2}} e^{-i \pi \nu}\Gamma\big(\nu - \frac{L d}{2}\big)}{\prod_{i=1}^N \Gamma(\nu_i)} \Bigg[ \prod_{j=1}^N \int_0^{\infty} {\rm d} x_j \Bigg] \delta(1-x_N)\,\mathcal{U}_{\rm M}^{\nu - (L+1)d/2} \mathcal{F}_{\rm M}^{L d/2-\nu}\prod_{k=1}^N x_k^{\nu_k-1},
\end{align}
\end{changemargin}
where $\mathcal{U}_{\rm M}$ and $\mathcal{F}_{\rm M}$ are the first and second Symanzik polynomials in Minkowski space. For our subsequent analysis of the Baikov-Lee representation, it is important to note that the functional dependence of $I_{\rm E}$ and $I_{\rm M}$ on the external kinematics is nearly identical.\footnote{At this juncture, it is of critical importance to clarify that, strictly speaking, this statement is not true. Obviously, a vector in Euclidean space which squares to zero is identically zero, whereas this is not the case in Minkowski space. However, one may simply write a formal expression for a Euclidean Feynman integral with the squares of certain momenta set to zero, remembering that, to be rigorous, one would have to work out the connection between Euclidean and Minkowski space representations with fake external masses and then set them to zero after the fact.} Given some spanning set of external kinematic invariants, $\{\omega_1,\ldots,\omega_{n(n-1)/2}\}$, constructed along the lines described in \cite{Bern:1992em}, we can straightforwardly obtain one from the other,
\begin{align}
\label{eq:E-M-map}
I_{\rm M} =  i^L e^{-i \pi \nu} I_{\rm E}\bigg|_{\omega_i \to -\omega_i,\,\{p_j^{*}\}\to -\{p_j^{*}\}},
\end{align}
provided that we remember the $+ i 0$ prescription for the external kinematic invariants and define $\{p_j^{*}\}$ to be the set of external momenta which happen to be incoming in the physical kinematics of interest. To understand the above relation, recall that $\mathcal{U}_{\rm M} = \mathcal{U}_{\rm E}$ and that one can generate the Minkowski space function $\mathcal{F}_{\rm M}$ from $\mathcal{F}_{\rm E}$ by flipping the signs of all generalized Mandelstam variables and external masses which appear and, subsequently, appropriately adjusting the signs of the external momenta which must now be regarded as incoming.\footnote{Naturally, we have assumed that Euclidean and Minkowskian generalized Mandelstam invariants are defined in the usual way. For instance, one would have $t = (p_1 + p_3)^2$ in Euclidean space but $t = (p_1 - p_3)^2$ in the usual physical kinematics for $2 \to 2$ scattering.}

For what concerns the explicit examples discussed in the following sections, we essentially adopt the conventions of reference~\cite{vonManteuffel:2014qoa}. That is to say, for our actual calculations, we consider Minkowskian purely-virtual Feynman integrals of the form
\begin{align}
\label{eq:gen-virt-int}
  I = \int\!\frac{{\rm d}^d k_1}{i \pi^{d/2}}  \cdots \int\!\frac{{\rm d}^d k_L}{i \pi^{d/2}} ~\prod_{\ell = 1}^N \left(Q_\ell^2(k_i,p_j) - m_\ell^2 + i 0\right)^{-\nu_\ell}
\end{align}
in physical kinematics. Note that we do not include factors in the measure to prevent the Euler-Mascheroni constant from appearing in $\ep$-expanded expressions because, in this paper, we either study cut Feynman integrals at $\mathcal{O}(\ep^0)$ or to all orders in $\ep$. To simplify our discussion later on, it is also convenient to introduce complex-conjugated purely-virtual Feynman integrals, $I^\dagger$, where the $+i 0$ prescription becomes a $-i 0$ prescription and the $i \pi^{d/2}$ factors in (\ref{eq:gen-virt-int}) above are replaced by factors of $-i \pi^{d/2}$. The maximally-cut examples of Section \ref{sec:maxcut} are far less sensitive to such details because overall phases make no difference at all if the only goal is to produce a valid solution to a given homogeneous differential equation. For the sake of definiteness, we will use the same normalization conventions in both Sections \ref{sec:stdcuts} and \ref{sec:maxcut}.


\subsection{The Euclidean Baikov-Lee representation and its analytical continuation}
\label{subsec:B-L-continuation}
To write the Euclidean Baikov-Lee formula succinctly, let us first recall that the Gram determinant on the $K$ linearly independent vectors $\{q_i\}$ is given by
\begin{align}
  G(q_1,\ldots,q_K) = \begin{vmatrix}
    q_1^2  &  \cdots  &  ~q_1\cdot q_K \\ \\
    \vdots  &  \ddots  &  \vdots \\ \\
    ~q_1\cdot q_K~  &  \ldots  &  q_K^2
  \end{vmatrix}.
\end{align}
If we let $q_i$ be an element of the set $\{k_1,\ldots,k_L,P_1,\ldots,P_{n-1}\}$,\footnote{Here, $\{P_i\}$ is nothing but a convenient permutation of the set of independent external momenta, $\{p_i\}$. This formulation is convenient because it is often desirable to eliminate a momentum other than the $n$-th. Our treatment of the $s$-cut of the massless one-loop box integral in Section \ref{subsec:1masslessbox} clearly illustrates this point.} the Baikov-Lee representation of the purely-virtual Feynman integral $I_{\rm E}$ defined above in Eq. (\ref{eq:gen-virt-int-E}) is then
\begin{align}
\label{eq:leerep-E}
  I_{\rm E} &= \frac{\pi^{L(3+2 d-2 n-L)/4}}{\prod_{r=0}^{L-1} \Gamma\left(\frac{d-n-r+1}{2}\right)\left[G(P_1,\ldots,P_{n-1})\right]^{(d-n)/2}}~\underset{\mathcal{D}}{\int\cdots\int} \left( \prod_{i=1}^L \prod_{j=i}^{n+L-1} {\rm d}(q_i \cdot q_j)\right) \times 
  \nonumber \\
  & \qquad \qquad \times \left[G(q_1,\ldots,q_{n + L - 1})\right]^{(d-n-L)/2}\prod_{\ell = 1}^N \left(Q_\ell^2(q_i\cdot q_j) + m_\ell^2\right)^{-\nu_\ell},
\end{align}
where $\mathcal{D}$ is the domain of integration. 

Even if one works in Euclidean space, finding an explicit description of the integration domain is in general a non-trivial task. To understand how this works in practice, let us consider the evaluation of the one-loop bubble with no internal masses in the Baikov-Lee approach. This example will both illustrate a general strategy for the determination of the integration region (briefly discussed in~\cite{Frellesvig:2017aai}) and give the reader a sense as to why it is more convenient to integrate purely-virtual Feynman integrals using Feynman parameters as a starting point. The arguments advanced in this section are generally applicable, but it may be quite challenging to work out the details in examples with many integration variables and/or rich analytic structures. For future applications, we expect tools for the explicit solution of systems of inequalities such as the {\tt Reduce} routine of {\tt Mathematica} to play an important role.

For the one-loop bubble, a possible routing of the propagator momenta is
\begin{align}
  Q_1 = k_1 \qquad \qquad \qquad Q_2 = k_1 - p\,. \nonumber
\end{align}
In this case, (\ref{eq:leerep-E}) becomes
\begin{align}
  \Graph{.3}{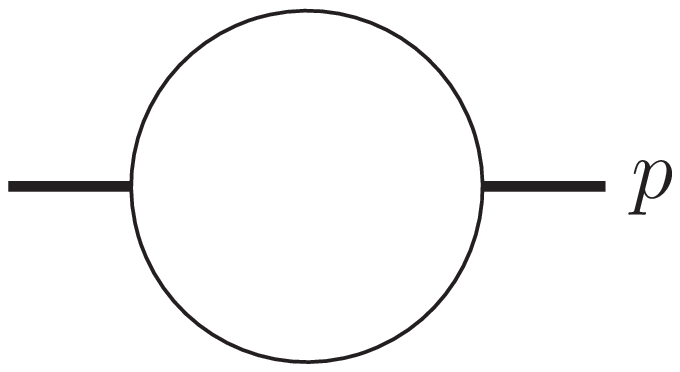} = \underset{\mathcal{D}}{\int}\int {\rm d}(q_1^2){\rm d}(q_1 \cdot q_2)\frac{\pi^{\frac{3}{2}-\ep}\left(p^2 q_1^2-(q_1\cdot q_2)^2\right)^{\frac{1}{2}-\ep}}{\Gamma\left(\frac{3}{2}-\ep\right)\left(p^2\right)^{1-\ep} q_1^2\left(q_1^2-2 q_1 \cdot q_2 + p^2\right)},
\end{align}
where we have set $d = 4 - 2\ep$. At this stage, it is important to note that the form of Eq. (\ref{eq:gen-parametric-rep-E}) and the definitions of $\mathcal{U}_{\rm E}$ and $\mathcal{F}_{\rm E}$ (see {\it e.g.} \cite{Bogner:2010kv}) together imply that Euclidean Feynman integrals are positive definite if all input kinematic variables are positive definite. This is a powerful analytic constraint from the Baikov-Lee point of view and effectively determines the shape of the integration region. The Baikov polynomial $p^2 q_1^2-(q_1\cdot q_2)^2$ inside the integrand above depends on the variables of integration and is raised to a non-integer power. This is a generic feature of Baikov-Lee calculations. The point is that, in order to keep the solution real-valued, one must consider {\it e.g.} $q_1\cdot q_2$ to lie between the branching points $\pm \sqrt{p^2 q_1^2}$. 

Apart from the obvious positivity of $q_1^2$, there are no further constraints on the variables of integration in this case, and we arrive at
\begin{align}
\label{eq:one-loop-bubble-explicit}
  \Graph{.3}{virtoneloopbubble} = \int_0^{\infty} {\rm d}(q_1^2)\int_{-\sqrt{p^2 q_1^2}}^{\sqrt{p^2 q_1^2}}{\rm d}(q_1 \cdot q_2)\frac{\pi^{\frac{3}{2}-\ep}\left(p^2 q_1^2-(q_1\cdot q_2)^2\right)^{\frac{1}{2}-\ep}}{\Gamma\left(\frac{3}{2}-\ep\right)\left(p^2\right)^{1-\ep} q_1^2\left(q_1^2-2 q_1 \cdot q_2 + p^2\right)}.
\end{align}
Although, the Baikov polynomial is no more than quadratic in the scalar products involving the loop momenta, the situation may become more complicated once the first scalar product is integrated out. In favorable cases, it is possible to find an analogous integration variable at each step of the calculation. However, it is not guaranteed that all polynomial structures remaining in the integrand after some number of integration steps have at most quadratic dependence on the remaining variables of integration. Although it is probably clear already, let us emphasize that the two-fold integral above is far, far more complicated than the trivial one-fold integral which one finds in the Feynman parametric approach to this simple problem. The evaluation of (\ref{eq:one-loop-bubble-explicit}) involves non-trivial hypergeometric function identities, and we refer the interested reader to Appendix \ref{app:Baikov} for a detailed discussion.

As we shall see, the procedure described above for the limits of integration is conceptually even simpler for Feynman integrals cut in a single kinematic channel. In such cases, one can also make use of the fact that the region of integration is bounded; cut Feynman integrals of this type will be closely related to the real radiative master integrals for some physical decay or scattering process which has a finite amount of energy and momentum in the initial state~\cite{Gehrmann-DeRidder:2003pne}. The story is otherwise analogous to what was described above for purely-virtual Euclidean Feynman integrals because, up to phase, one again has a natural positivity condition. That is to say, in favorable cases, one can integrate each variable between branching points of the current integrand and then assign a positive orientation to the integration contour for the current variable ({\it i.e.} one must integrate $q_1 \cdot q_2$ from $-\sqrt{p^2 q_1^2}$ to $\sqrt{p^2 q_1^2}$ in the above example, not vice versa). We have applied these ideas to explicitly evaluate a variety of Feynman integrals cut in a single kinematic channel at one, two, and three loops.

Of course, before defining the cut Baikov-Lee representation which we will study throughout the rest of this paper, we first need to analytically continue Eq. (\ref{eq:leerep-E}). The key idea is to recognize that (\ref{eq:gen-parametric-rep-E}) and (\ref{eq:leerep-E}) are nothing but two different integral representations of the same function, $I_{\rm E}$. Since our recipe to pass from $I_{\rm E}$ to $I_{\rm M}$, Eq. (\ref{eq:E-M-map}), does not depend at all on the details of the Feynman representation, it is natural to apply it to Eq. (\ref{eq:leerep-E}) as well,\footnote{Due to the fact that the Baikov-Lee representation utilizes scalar product integration variables which have non-trivial dependence on the external momenta, it is not obvious that one can work in this way. However, we have found experimentally that this prescription does in fact make sense provided that one allows $\{P_j^{*}\}\to -\{P_j^{*}\}$ to act on the relevant scalar product integration variables as well.} thereby obtaining a putative physical, Minkowski space version of the Baikov-Lee representation,
\begin{align}
\label{eq:leerep-M}
  I_{\rm M} &= \frac{i^L \pi^{L(3+2 d-2 n-L)/4}e^{-i \pi \nu}}{\prod_{r=0}^{L-1} \Gamma\left(\frac{d-n-r+1}{2}\right)\left[G(P_1,\ldots,P_{n-1})\right]^{(d-n)/2}}~\underset{\mathcal{D}}{\int\cdots\int} \left( \prod_{i=1}^L \prod_{j=i}^{n+L-1} {\rm d}(q_i \cdot q_j)\right) \times 
  \nonumber \\
  & \qquad \quad \times \left[G(q_1,\ldots,q_{n + L - 1})\right]^{(d-n-L)/2} \prod_{\ell = 1}^N \left(Q_\ell^2(q_i\cdot q_j) + m_\ell^2\right)^{-\nu_\ell}\bigg|_{\omega_i \to -\omega_i,\,\{P_j^{*}\}\to -\{P_j^{*}\}} 
  \nonumber \\
  &= \frac{i^L \pi^{L(3+2 d-2 n-L)/4}e^{-i \pi \nu}}{\prod_{r=0}^{L-1} \Gamma\left(\frac{d-n-r+1}{2}\right)\left[\bar{G}(P_1,\ldots,P_{n-1})\right]^{(d-n)/2}}~\underset{\bar{\mathcal{D}}}{\int\cdots\int} \left( \prod_{i=1}^L \prod_{j=i}^{n+L-1} {\rm d}(q_i \cdot q_j)\right) \times 
  \nonumber \\
  & \qquad \quad \times \left[\bar{G}(q_1,\ldots,q_{n + L - 1})\right]^{(d-n-L)/2}\prod_{\ell = 1}^N \left(\bar{Q}_\ell^2(q_i\cdot q_j) + m_\ell^2\right)^{-\nu_\ell},
\end{align}
where $\bar{G}(P_1,\ldots,P_{n-1})$, $\bar{\mathcal{D}}$, $\bar{G}(q_1,\ldots,q_{n + L - 1})$, and $\bar{Q}_\ell(q_i\cdot q_j)$ denote the various objects which appear in Eq. (\ref{eq:leerep-M}) after the replacements prescribed by (\ref{eq:E-M-map}) have been implemented.


\subsection{The cut Baikov-Lee representation and unitarity}
The key ingredient missing from the discussion so far is the generalized Cutkosky cutting rule written down by Lee and Smirnov in reference~\cite{Lee:2012te}. In a nutshell, they suggest that one can treat cut Feynman integrals with propagator denominators raised to powers greater than one by simply differentiating both sides of Cutkosky's relation,
\begin{align}
  \frac{1}{k^2+i0} - \frac{1}{ k^2 - i0} = -2\pi i \theta(k^0) \delta\left(k^2\right),
\end{align}
an appropriate number of times with respect to $k^2$.\footnote{We thank Gil Paz for pointing out that the distributional identity behind the generalized cutting rule was available in textbooks on the subject ({\it e.g.}~\cite{GelfandShilov:1964}) long before the appearance of reference~\cite{Lee:2012te}.} To avoid digressing into a lengthy discussion of distributional calculus,\footnote{The direct integration of delta distributions and their derivatives is straightforward~\cite{Folland:1992}, but it seems somewhat less convenient from the perspective of implementation in a computer algebra system.} it is convenient to actually define our Baikov-Lee representation for Feynman integrals cut in a single kinematic channel using the familiar language of residue calculus. The idea is that, up to a possible overall sign, the process of putting some number of propagators on the mass shell is completely equivalent to performing sequential residue computations which localize a subset of the scalar product integration variables. Our logic is similar to that of reference~\cite{Abreu:2017ptx}, except that, for our purposes, we find it more natural to repeatedly apply Cauchy's residue theorem to Eq. (\ref{eq:leerep-M}). Due to the fact that we will use the main result of this section for the explicit examples discussed in the following sections, we find it natural to work with the absolute normalization of Eq. (\ref{eq:gen-virt-int}) in what follows.

For the sake of discussion, suppose that a particular unitarity cut of $I$ in the $\omega_i$ channel, say the $j$-th out of $M$, puts $n_j$ propagators on shell. By assumption, these propagators are linearly independent and there must therefore exist a subset of the scalar products depending on the loop momenta which one can sequentially integrate out using $n_j$ applications of the residue theorem. To simplify our notation, let $\{\bar{s}_i\}$ be the subset of the scalar product integration variables to be localized by the cut propagators, $\{s_k\}$ be the set of $L(L-1)/2 + n L - n_j$ scalar product integration variables left over,\footnote{In some cases, such as that of the one-loop double-cut bubble integral, the set $\{s_k\}$ is actually empty.} and $\{\bar{Q}_\ell\}^{(j)}_{\omega_i{\rm -cut}}$ be the momenta of the cut propagators. Note that, at this stage, the order of both $\{\bar{s}_i\}$ and $\{\bar{Q}_\ell\}^{(j)}_{\omega_i{\rm -cut}}$ should be fixed to match the order in which the localization of the propagators and associated scalar products will be implemented by the residue theorem. 

Finally, we define the Baikov-Lee representation of the $j$-th Feynman integral cut in the $\omega_i$ channel to be
\begin{align}
\label{eq:leerepcut}
  &I_{\omega_i{\rm -cut}}^{(j)} = \frac{(-2 \pi i)^{n_j} \pi^{L(3-2 n-L)/4}e^{-i \pi \nu}}{\prod_{r=0}^{L-1} \Gamma\left(\frac{d-n-r+1}{2}\right)\left[\bar{G}(P_1,\ldots,P_{n-1})\right]^{(d-n)/2}}~\underset{\bar{\mathcal{D}}^{(j)}_{\omega_i{\rm -cut}}}{\int\cdots\int} \left( \prod_{k=1}^{L(L-1)/2 + n L - n_j} {\rm d}s_k\right) \times 
  \nonumber \\
  & \times \mathbf{sgn}\left(\left|\frac{\partial \{\bar{Q}^2_\ell\}^{(j)}_{\omega_i{\rm -cut}}}{\partial \{\bar{s}_i\}}\right|\right) \underset{{\{\bar{s}_i\}}}{\mathbf{Res}}\left\{ \left[\bar{G}(q_1,\ldots,q_{n + L - 1})\right]^{(d-n-L)/2} \prod_{\ell = 1}^N \left(\bar{Q}_\ell^2(\bar{s}_i,s_k) + m_\ell^2\right)^{-\nu_\ell}\right\},
\end{align}
where the $\mathbf{sgn}$ function returns the sign of its argument and $\mathbf{Res}$ denotes the sequence of residue computations which localizes the $\{\bar{s}_i\}$. The presence of the $\mathbf{sgn}$ factor is necessary because sequential residue computations do differ from sequential localizations implemented with delta distributions and their derivatives in one important aspect. The issue is that
\begin{align}
\underset{{\{a\}}}{\mathbf{Res}} \left\{ \frac{f(z)}{a-z} \right\} = - f(a),
\end{align}
but
\begin{align}
\int_{-\infty}^\infty {\rm d}z \,\delta(a-z)f(z) = \int_{-\infty}^\infty {\rm d}z \,\delta(z-a)f(z) = f(a)
\end{align}
for arbitrary test functions $f(z)$ regular at $z = a$. In fact, one must include the sign of the Jacobian factor in (\ref{eq:leerepcut}) above, or the definition yields nonsensical results which may vary depending upon precisely what momentum routing is chosen for the cut Feynman integral under consideration. 

We explicitly evaluate a number of cut Feynman integrals in Section \ref{sec:stdcuts} using Eq. (\ref{eq:leerepcut}) as a starting point, and we find that our definition is consistent in all cases.
In particular, we find the expected unitarity relation between sums of cut Feynman integrals and the direct discontinuities\footnote{We define the direct discontinuity of a Feynman integral in Appendix \ref{app:virtual}.} of their purely-virtual counterparts~\cite{tHooft:1973wag},
\begin{align}
\label{eq:unitarity-easy}
  {\rm Disc}_{\omega_i}\left(I\right)  = -\sum_{j = 1}^M I_{\omega_i{\rm -cut}}^{(j)}.
\end{align}
In fact, we have successfully used our framework to study a large number of other examples of comparable complexity at one, two, and three loops. However, since we have no proof that our formulation is equivalent to the usual one where one considers all Feynman integrals to be embedded in an ambient generalized scalar field theory, it is important to write down a cut integral definition and associated unitarity relation along the lines of \cite{Gehrmann-DeRidder:2003pne,Abreu:2014cla}. If $\hat{I} = i^L I$, then
\begin{align}
\label{eq:leerepcut-trad}
  &\hat{I}_{\omega_i{\rm -cut}}^{(j)} = \frac{i^L (2 \pi)^{n_j} \pi^{L(3-2 n-L)/4}e^{-i \pi \nu}}{\prod_{r=0}^{L-1} \Gamma\left(\frac{d-n-r+1}{2}\right)\left[\bar{G}(P_1,\ldots,P_{n-1})\right]^{(d-n)/2}}~\underset{\bar{\mathcal{D}}^{(j)}_{\omega_i{\rm -cut}}}{\int\cdots\int} \left( \prod_{k=1}^{L(L-1)/2 + n L - n_j} {\rm d}s_k\right) \times 
  \nonumber \\
  & \times \mathbf{sgn}\left(\left|\frac{\partial \{\bar{Q}^2_\ell\}^{(j)}_{\omega_i{\rm -cut}}}{\partial \{\bar{s}_i\}}\right|\right) \underset{{\{\bar{s}_i\}}}{\mathbf{Res}}\left\{ \left[\bar{G}(q_1,\ldots,q_{n + L - 1})\right]^{(d-n-L)/2} \prod_{\ell = 1}^N \left(\bar{Q}_\ell^2(\bar{s}_i,s_k) + m_\ell^2\right)^{-\nu_\ell}\right\}
\end{align}
and
\begin{align}
\label{eq:unitarity-hard}
{\rm Disc}_{\omega_i}\left(\mathcal{P}\left\{\hat{I}\right\} \hat{I}\right) = \sum_{j = 1}^M \mathcal{P}\left\{\hat{I}_{\omega_i{\rm -cut}}^{(j)}\right\} \hat{I}_{\omega_i{\rm -cut}}^{(j)},
\end{align}
where $\mathcal{P}\left\{\hat{I}\right\}$ and $\mathcal{P}\left\{\hat{I}_{\omega_i{\rm -cut}}^{(j)}\right\}$ are scalar field theory phase factors defined in {\it e.g.}~\cite{Abreu:2014cla}. 

For the purposes of this paper, we employ Eq. (\ref{eq:leerepcut}) with the attitude that it streamlines the exposition in Section \ref{sec:stdcuts} and makes it easier for the reader to check our analysis using the results from the literature collected in Appendix \ref{app:virtual} ({\it i.e.} it allows us to forget about the annoying Feynman graph-dependent phase factors on both sides of (\ref{eq:unitarity-hard})). Before leaving this section, let us emphasize that one can also employ the formalism discussed above to treat maximally-cut Feynman integrals in an analogous fashion. Interesting examples of maximally-cut Feynman integrals will be discussed in Section \ref{sec:maxcut}.


\section{Discontinuities from cuts: one- and two-loop examples}
\label{sec:stdcuts}
To get a feeling for the ideas put forward in Section \ref{sec:genform}, we now consider a number of illustrative one- and two-loop examples. In the spirit of reference \cite{Abreu:2014cla}, we compute the $s$-channel cuts of selected Feynman integrals using our cut Baikov-Lee representation, Eq. (\ref{eq:leerepcut}), and demonstrate that, in all cases, our results match the predictions of the optical theorem ({\it i.e.} the predictions obtained by using Appendix \ref{app:virtual} to compute the direct discontinuities on the left-hand side of Eq. (\ref{eq:unitarity-easy})).


\subsection{The one-external-mass one-loop triangle}
\label{subsec:triangle-gendots}
In this section, we consider the $s$-channel cut of the one-external-mass one-loop triangle with positive integer propagator exponents,
\begin{align}
\label{eq:triangle-gendots-def}
  \Graph{.3}{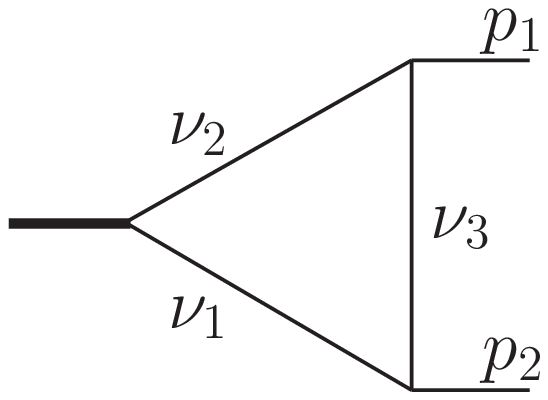} = \int \frac{d^d k_1}{i \pi^{d/2}}~ \frac{1}{[(p_2-k_1)^2]^{\nu_1} [(p_1+k_1)^2]^{\nu_2} [k_1^2]^{\nu_3}}.
\end{align}
This example clearly demonstrates the applicability of our formalism to propagators of higher multiplicity, and its elementary nature should give the reader ample opportunity to adjust to our notation. Clearly, there is only one cut Feynman integral which needs to be evaluated. We have
\begin{align}
\label{eq:cuttriangle-Baikov-def}
  \Graph{.3}{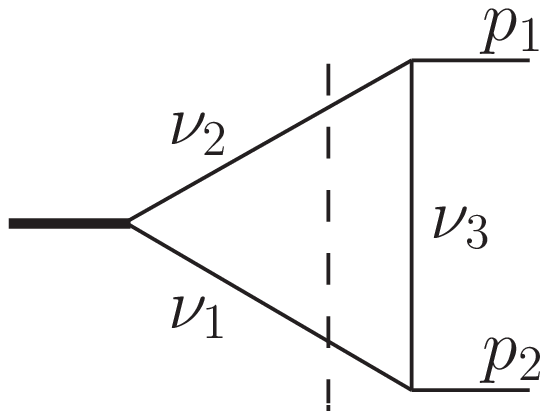} &= \frac{(-2\pi i)^2 \pi^{-1} (-1)^{\nu}}{\Gamma (1-\ep)\left(-s^2/4\right)^{1/2-\ep}} \mathbf{sgn}\left(\left|\frac{\partial \{s_1-2 \bar{s}_1,s_1+2 \bar{s}_2\}}{\partial \{\bar{s}_1,\bar{s}_2\}}\right|\right)\times
  \nonumber \\
  &\times \underset{\bar{\mathcal{D}}_{s{\rm -cut}}}{\int} {\rm d}s_{1}~\underset{{\{\bar{s}_1,\bar{s}_2\}}}{\mathbf{Res}}\left\{\frac{\left(-\bar{s}_1 \bar{s}_2 s-s_1 s^2/4\right)^{-\ep }}{\left(s_1-2 \bar{s}_1\right)^{\nu_1}\left(s_1+2\bar{s}_2\right)^{\nu_2} s_1^{\nu_3}}\right\}
\end{align}
in the cut Baikov-Lee representation of Eq. (\ref{eq:leerepcut}), where $s_1 = k_1^2$, $\bar{s}_1 = k_1 \cdot p_2$, $\bar{s}_2 = k_1 \cdot p_1$, and $s = (p_1 + p_2)^2$.

Eq. (\ref{eq:cuttriangle-Baikov-def}) can be conveniently rewritten as
\begin{align}
\label{eq:cuttriangle-Baikov-simp}
  \Graph{.3}{cutonelooptriangle} &= \frac{2^{3-\nu_1-\nu_2} \pi (-1)^{\nu_2+\nu_3}i}{s^{1-\ep}\Gamma (1-\ep)} \times
  \nonumber \\
  &\times \underset{\bar{\mathcal{D}}_{s{\rm -cut}}}{\int} \frac{{\rm d}s_{1}}{s_1^{\nu_3}}~\underset{{\{\bar{s}_1,\bar{s}_2\}}}{\mathbf{Res}}\left\{\frac{\left(4\bar{s}_1 \bar{s}_2+s_1 s\right)^{-\ep }}{\left(\bar{s}_1-s_1/2\right)^{\nu_1}\left(\bar{s}_2+s_1/2\right)^{\nu_2}}\right\}
\end{align}
to make manifest the fact that its right-hand side is purely imaginary for $s > 0$ and to streamline the applications of the residue theorem which follow. Using the principle of mathematical induction, it is straightforward to evaluate the symbolic derivatives which enter into the residue calculations. Reading the list of barred integration variables from left to right, we find
\begin{align}
\label{eq:cuttriangle-Baikov-res1}
  \Graph{.3}{cutonelooptriangle} &= -\frac{2^{1+\nu_1-\nu_2} \pi \Gamma(\nu_1-1+\ep) (-1)^\nu i}{s^{1-\ep}\Gamma(\ep)\Gamma (1-\ep)\Gamma(\nu_1)} \times
  \nonumber \\
  &\times \underset{\bar{\mathcal{D}}_{s{\rm -cut}}}{\int} \frac{{\rm d}s_{1}}{s_1^{\nu_1+\nu_3-1+\ep}}~
  \underset{{\{\bar{s}_2\}}}{\mathbf{Res}}\left\{\frac{\bar{s}_2^{\nu_1-1}\left(s+2\bar{s}_2\right)^{1-\nu_1-\ep }}{\left(\bar{s}_2+s_1/2\right)^{\nu_2}}\right\}
  \nonumber \\
  &= -\frac{2\pi \Gamma(\nu_1-1+\ep) (-1)^{\nu_3} i}{s^{2-\nu_2-\ep}\Gamma(\ep)\Gamma (1-\ep)\Gamma(\nu_2)\Gamma(1+\nu_1-\nu_2)}\underset{\bar{\mathcal{D}}_{s{\rm -cut}}}{\int} \frac{{\rm d}s_{1}}{s_1^{\nu_2+\nu_3-1+\ep}}\times
  \nonumber \\
  &\times (s-s_1)^{2-\nu_1-\nu_2-\ep} {}_2F_1\left(1-\nu_2,2-\nu_2-\ep;1+\nu_1-\nu_2;\frac{s_1}{s}\right).
\end{align}

At this stage of the calculation, we have to analyze the integrand to determine the $s_1$ integration domain. By studying the form of Eq. (\ref{eq:cuttriangle-Baikov-res1}), we see immediately that the integrand has precisely two branching points, at $s_1 = 0$ and at $s_1 = s$. From the general discussion in Section \ref{subsec:B-L-continuation}, it therefore follows that
\begin{align}
\label{eq:cuttriangle-Baikov-res2}
  \Graph{.3}{cutonelooptriangle} &= -\frac{2\pi \Gamma(\nu_1-1+\ep) (-1)^{\nu_3} i}{s^{2-\nu_2-\ep}\Gamma(\ep)\Gamma (1-\ep)\Gamma(\nu_2)\Gamma(1+\nu_1-\nu_2)}\int_0^s \frac{{\rm d}s_{1}}{s_1^{\nu_2+\nu_3-1+\ep}}\times
  \nonumber \\
  &\times (s-s_1)^{2-\nu_1-\nu_2-\ep} {}_2F_1\left(1-\nu_2,2-\nu_2-\ep;1+\nu_1-\nu_2;\frac{s_1}{s}\right).
\end{align}
For $\ep$ such that $\mathfrak{Re}(\nu_1+\nu_2+\ep) < 3$ and $\mathfrak{Re}(\nu_2+\nu_3+\ep) < 2$, the above integral converges and we find that
\begin{align}
\label{eq:cuttriangle-res}
  &\Graph{.3}{cutonelooptriangle} = -\frac{2\pi s^{2-\nu-\ep}\Gamma(\nu_1-1+\ep)\Gamma(3-\nu_1-\nu_2-\ep)\Gamma(2-\nu_2-\nu_3-\ep)(-1)^{\nu_3}i}{\Gamma(\ep)\Gamma(1-\ep)\Gamma(\nu_2)\Gamma(1+\nu_1-\nu_2)\Gamma(5-\nu_1-2\nu_2-\nu_3-2\ep)}\times 
  \nonumber\\
  &\times {}_3F_2(1-\nu_2,2-\nu_2-\ep,2-\nu_2-\nu_3-\ep;1+\nu_1-\nu_2,5-\nu_1-2\nu_2-\nu_3-2\ep;1).
\end{align}

Under an additional assumption, Eq. (\ref{eq:cuttriangle-res}) can be simplified using the Saalsch{\"u}tz summation formula. For $\nu_2 > 1$, Eq. (\ref{eq:Saalschutz}) implies that
\begin{align}
\label{eq:one-loop-triangle-cut}
  \Graph{.3}{cutonelooptriangle} = -\frac{2 i \sin(\pi \ep) s^{2-\nu-\ep} \Gamma(2-\nu_1-\nu_3-\ep) \Gamma(2-\nu_2-\nu_3-\ep) \Gamma(\nu-2+\ep)}{\Gamma(\nu_1) \Gamma(\nu_2) \Gamma(4-\nu-2\ep)}.
\end{align}
Actually, the principle of analytical continuation allows us to conclude that Eq. (\ref{eq:one-loop-triangle-cut}) is not only valid for arbitrary positive integer propagator exponents as desired, but that it even furnishes a definition of the cut Feynman integral for generic complex values of the propagator exponents. In fact, this analytical continuation of the above result will prove useful later on in Section \ref{subsec:2formfac}. Finally, one can readily check using (\ref{eq:one-loop-triangle-cut}) and the $s > 0$ evaluation of (\ref{eq:triangle-gendots-def}) given in Eq. (\ref{eq:one-loop-triangle-virt}) that
\begin{align}
{\rm Disc}_{s}\left(\Graph{.3}{virtonelooptriangle}\right)  = -\left(\Graph{.3}{cutonelooptriangle}\right).
\end{align}


\subsection{The massless one-loop box}
\label{subsec:1masslessbox}
For our next example, we consider the $s$-channel cut of the massless one-loop box integral,
\begin{align}
\label{eq:box-def}
  \Graph{.3}{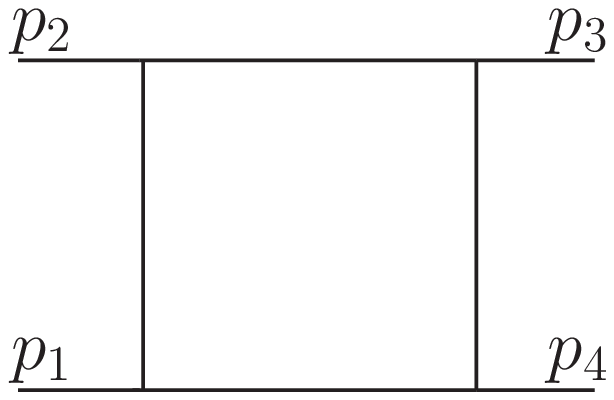} = \int \frac{d^d k_1}{i \pi^{d/2}}~ \frac{1}{k_1^2 (k_1+p_1)^2 (k_1+p_1+p_2)^2 (k_1+p_4)^2}.    
\end{align}
Although it is again the case that just one cut Feynman integral needs to be evaluated, it is useful to study this example because it illustrates the applicability of our framework to multi-scale problems. In fact, it is not obvious to us that this cut can be computed to all orders in $\ep$ using traditional cut parameterizations. We shall see that it is also not entirely straightforward in the Baikov-Lee approach; the cut Feynman integral of interest here contains structures which bear a remarkable resemblance to those which appeared during the evaluation of certain three-loop, single-scale cut Feynman integrals~\cite{Li:2014bfa}.

From Eq. (\ref{eq:leerepcut}), we obtain
\begin{align}
\label{eq:box-cut-BL}
  &\Graph{.3}{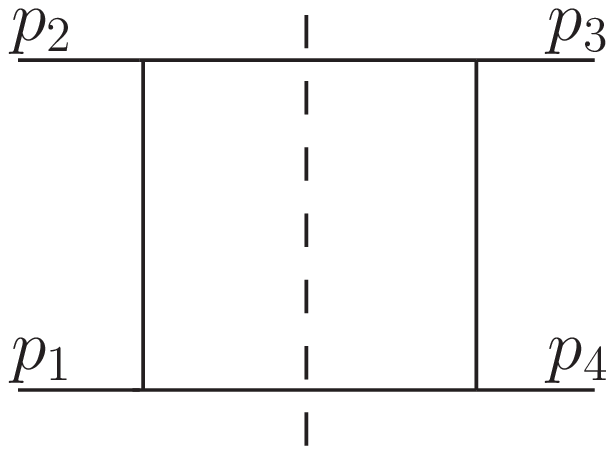} = -\frac{4 \pi^{1/2} (-s u (s + u))^{\ep} i}{\Gamma (1/2-\ep)} \underset{\bar{\mathcal{D}}_{s{\rm -cut}}}{\int \int} {\rm d}s_1{\rm d}s_2\times
  \\
  &\times \underset{{\{\bar{s}_1,\bar{s}_2\}}}{\mathbf{Res}}\left\{\frac{\Big(-\bar{s}_1 s u (s+u) - s_1^2 (s+u)^2 - (\bar{s}_2 u + s_2 s)^2 - 2 s_1 (s+u)(\bar{s}_2 u - s_2 s)\Big)^{-1/2-\ep}}{\bar{s}_1\left(\bar{s}_1+2s_1\right)\left(\bar{s}_2+s_1-s/2+\bar{s}_1/2\right)\left(\bar{s}_1+2 s_2\right)}\right\} 
  \nonumber
\end{align}
for our cut Baikov-Lee representation after performing some trivial manipulations to massage the expressions into a more convenient form. In Eq. (\ref{eq:box-cut-BL}), we have made the definitions $s_1 = k_1 \cdot p_1$, $s_2 = k_1 \cdot p_4$, $\bar{s}_1 = k_1^2$, $\bar{s}_2 = k_1 \cdot p_2$, $s = (p_1 + p_2)^2$, and $u = (p_1 - p_4)^2$. Carrying out the residue computations, we find
\begin{align}
\label{eq:box-cut-res}
  \Graph{.3}{cutoneloopbox} &= -\frac{\pi^{1/2} (-s u (s + u))^{\ep} i}{\Gamma (1/2-\ep)} \underset{\bar{\mathcal{D}}_{s{\rm -cut}}}{\int \int} \frac{{\rm d}s_1}{s_1}\frac{{\rm d}s_2}{s_2}\times
  \nonumber \\
  &\times \underset{{\{\bar{s}_2\}}}{\mathbf{Res}}
  \left\{\frac{\Big(- s_1^2 (s+u)^2 - (\bar{s}_2 u + s_2 s)^2 - 2 s_1 (s+u)(\bar{s}_2 u - s_2 s)\Big)^{-1/2-\ep}}{\bar{s}_2+s_1-s/2}\right\} 
  \nonumber \\
  &= -\frac{2^{1+2\ep}\pi^{1/2} (-u (s + u))^{\ep} i}{s^{1/2} \Gamma (1/2-\ep)} \underset{\bar{\mathcal{D}}_{s{\rm -cut}}}{\int \int} \frac{{\rm d}s_1}{s_1}\frac{{\rm d}s_2}{s_2}\times
  \nonumber \\
  &\times \Big(16 s_1 s_2 u - s \left(u^2+4 (s_1+s_2) u+4 (s_1-s_2)^2\right)\Big)^{-1/2-\ep}.
\end{align}

Next, we integrate out the variable $s_2$. As explained in Section \ref{subsec:B-L-continuation}, the integration runs between the real zeros of the polynomial from the last line of Eq. (\ref{eq:box-cut-res}),
\begin{align}
s_2^{\pm} = \frac{s_1 (s +2 u) - s u/2 \pm \sqrt{-2 u (s+u)} \sqrt{s_1 (s-2 s_1)}}{s}.
\end{align}
In fact, by observing that the $s_2$ integration is completely analogous to the integration with respect to $q_1\cdot q_2$ carried out in Appendix \ref{app:Baikov}, we can already anticipate from the branch cut structure of the $s_2^{\pm}$ (with respect to $s_1$) that the $s_1$ integration will run between $0$ and $s/2$. We therefore readily obtain
\begin{align}
\label{eq:box-cut-int1}
  \Graph{.3}{cutoneloopbox} &= -\frac{2^{1+2\ep}\pi^{1/2} (-u (s + u))^{\ep} i}{s^{1/2} \Gamma (1/2-\ep)} \int_0^{s/2} \frac{{\rm d}s_1}{s_1} \int_{s_2^-}^{s_2^+} \frac{{\rm d}s_2}{s_2}\times
  \nonumber \\
  &\times \Big(16 s_1 s_2 u - s \left(u^2+4 (s_1+s_2) u+4 (s_1-s_2)^2\right)\Big)^{-1/2-\ep}
  \nonumber \\
  &= \frac{4\pi s^{-2-\ep} i}{u\Gamma(1-\ep)} \int_0^{s/2} {\rm d} s_1
  \frac{
  \left(\frac{2 s_1}{s}\right)^{-1-\ep}\left(1-\frac{2 s_1}{s}\right)^{-\ep}
  }{
  \left(\sqrt{\frac{s+u}{-u}\frac{2 s_1}{s}}+\sqrt{1-\frac{2 s_1}{s}}\right)^2
  }\times
  \nonumber \\
  &\times {}_2F_1\left(1,\frac{1}{2}-\ep;1-2\ep;
  \frac{
  4\sqrt{\frac{s+u}{-u}\frac{2 s_1}{s}\left(1-\frac{2 s_1}{s}\right)}
  }
  {
   \left(\sqrt{\frac{s+u}{-u}\frac{2 s_1}{s}}+\sqrt{1-\frac{2 s_1}{s}}\right)^2
  }\right)
  \nonumber \\
  &= \frac{2\pi s^{-1-\ep} i}{u\Gamma(1-\ep)} \int_0^1 {\rm d} x
  \frac{
  x^{-1-\ep}\left(1-x\right)^{-\ep}
  }{
  \left(\sqrt{\frac{s+u}{-u}x}+\sqrt{1-x}\right)^2
  }\times
  \nonumber \\
  &\times {}_2F_1\left(1,\frac{1}{2}-\ep;1-2\ep;
  \frac{
  4\sqrt{\frac{s+u}{-u}x\left(1-x\right)}
  }
  {
  \left(\sqrt{\frac{s+u}{-u}x}+\sqrt{1-x}\right)^2
  }\right),
\end{align}
where we have made the change of variables $s_1 = s/2\,x$ in the last line. 

At this point, the above result may be rewritten to exhibit a ${}_2F_1$ of argument $\frac{4 z}{(1+z)^2}$, where $z$ can be chosen to be either $\sqrt{\frac{1-x}{\frac{s+u}{-u}x}}$ or its reciprocal. From this observation, we see that a simple strategy to eliminate the square roots appearing in Eq. (\ref{eq:box-cut-int1}) is to split the integral at the point where $z = 1$ and then apply a quadratic hypergeometric function transformation which is valid for $|z|<1$ ({\it i.e.} Eq. (\ref{eq:2F1-quadratic})) to both terms. Carrying out these steps, we find
\begin{align}
\label{eq:box-cut-int-quad}
  \Graph{.3}{cutoneloopbox} &= \frac{2\pi s^{-1-\ep} i}{u\Gamma(1-\ep)} \int_0^{-u/s} {\rm d} x
  \frac{
  x^{-1-\ep}\left(1-x\right)^{-1-\ep}
  }{
  \left(1+\sqrt{\frac{\frac{s+u}{-u}x}{1-x}}\right)^2
  }\times
  \nonumber \\
  &\times {}_2F_1\left(1,\frac{1}{2}-\ep;1-2\ep;
  \frac{
  4\sqrt{\frac{\frac{s+u}{-u}x}{1-x}}
  }
  {
  \left(1+\sqrt{\frac{\frac{s+u}{-u}x}{1-x}}\right)^2
  }\right) 
  \nonumber \\
  &- \frac{2\pi s^{-1-\ep} i}{(s+u)\Gamma(1-\ep)} \int_{-u/s}^1 {\rm d} x
  \frac{
  x^{-2-\ep}\left(1-x\right)^{-\ep}
  }{
  \left(1+\sqrt{\frac{1-x}{\frac{s+u}{-u}x}}\right)^2
  }\times
  \nonumber \\
  &\times {}_2F_1\left(1,\frac{1}{2}-\ep;1-2\ep;
  \frac{
  4\sqrt{\frac{1-x}{\frac{s+u}{-u}x}}
  }
  {
  \left(1+\sqrt{\frac{1-x}{\frac{s+u}{-u}x}}\right)^2
  }\right)
  \\
  &= \frac{2\pi s^{-1-\ep} i}{u\Gamma(1-\ep)} \int_0^{-u/s} {\rm d} x~
  x^{-1-\ep}\left(1-x\right)^{-1-\ep} {}_2F_1\left(1,1+\ep;1-\ep;\frac{\frac{s+u}{-u}x}{1-x}\right)
  \nonumber \\
  &- \frac{2\pi s^{-1-\ep} i}{(s+u)\Gamma(1-\ep)} \int_{-u/s}^1 {\rm d} x~
  x^{-2-\ep}\left(1-x\right)^{-\ep} {}_2F_1\left(1,1+\ep;1-\ep;\frac{1-x}{\frac{s+u}{-u}x}\right). \nonumber
\end{align}

Now that the square root structures have been eliminated, we can deal with the two terms in (\ref{eq:box-cut-int-quad}) above by mapping them to linear combinations of known generalized Euler integrals. The first step is to make the change of variables $x = \frac{y}{\frac{s+u}{-u}+y}$ in the first integral and the change of variables $x = \frac{1}{1+\frac{s+u}{-u}y}$ in the second integral to bring them into the generalized Euler form. In fact, after making these transformations, one can immediately evaluate the second integral by applying integration formula (\ref{eq:2F1to3F2-int}):
\begin{align}
\label{eq:box-cut-res1}
  \Graph{.3}{cutoneloopbox} &= -\frac{2\pi s^{-1-\ep}(-u)^{-1-\ep} i}{(s+u)^{-\ep}\Gamma(1-\ep)} \int_0^1 {\rm d} y~
  y^{-1-\ep} \left(1-\frac{u}{s+u}y\right)^{2 \ep} {}_2F_1\left(1,1+\ep;1-\ep;y\right)
  \nonumber \\
  & - \frac{2\pi s^{-1-\ep}(-u)^{-1+\ep} i}{(s+u)^{\ep} \Gamma(1-\ep)} \int_0^1 {\rm d} y~
  y^{-\ep}\left(1-\frac{s+u}{u}y\right)^{2\ep} {}_2F_1\left(1,1+\ep;1-\ep;y\right)
  \nonumber \\
  &= -\frac{2\pi s^{-1-\ep}(-u)^{-1-\ep} i}{(s+u)^{-\ep}\Gamma(1-\ep)} \int_0^1 {\rm d} y~
  y^{-1-\ep} \left(1-\frac{u}{s+u}y\right)^{2 \ep} {}_2F_1\left(1,1+\ep;1-\ep;y\right)
  \nonumber \\
  &- \frac{2\pi \Gamma(-2\ep) s^{-1+\ep}(-u)^{-1-\ep} i}{(s+u)^{\ep} \Gamma(1-\ep) \Gamma(1-2\ep)}
  ~{}_3F_2\left(1,-2\ep,-2\ep;1-2\ep,1-\ep;1+\frac{u}{s}\right).
\end{align}

In our opinion, it is most convenient to deal with the remaining integral by replacing ${}_2F_1\left(1,1+\ep;1-\ep;y\right)$ with its integral representation, Eq. (\ref{eq:2F1-intdef}), and then applying Eq. (\ref{eq:F1-intdef}) to integrate out the variable $y$ with an Appell series. This yields
\begin{changemargin}{-.2 cm}{0 cm}
\begin{align}
\label{eq:box-cut-res2}
  \Graph{.3}{cutoneloopbox} &= -\frac{2\pi s^{-1-\ep}(-u)^{-1-\ep} i}{(s+u)^{-\ep}\Gamma(1-\ep)} \int_0^1 {\rm d} t \,(1-t)^{-1-\ep} F_1\left(-\ep;-2\ep,1+\ep;1-\ep;\frac{u}{s+u},t\right)
  \nonumber \\
  &- \frac{2\pi \Gamma(-2\ep) s^{-1+\ep}(-u)^{-1-\ep} i}{(s+u)^{\ep} \Gamma(1-\ep) \Gamma(1-2\ep)}
  ~{}_3F_2\left(1,-2\ep,-2\ep;1-2\ep,1-\ep;1+\frac{u}{s}\right).
\end{align}
\end{changemargin}
In this case, the sum of the second and third parameters of the $F_1$ is equal to the fourth parameter and reduction formula (\ref{eq:F1-reduction}) therefore immediately leads to
\begin{align}
\label{eq:box-cut-res3}
  \Graph{.3}{cutoneloopbox} &= -\frac{2\pi s^{-1-\ep}(-u)^{-1-\ep} i}{(s+u)^{-\ep}\Gamma(1-\ep)} \int_0^1 \frac{{\rm d} t}{1-t} \,{}_2F_1\left(-\ep,-2\ep;1-\ep;\frac{\frac{u}{s+u}-t}{1-t}\right)
  \nonumber \\
  &- \frac{2\pi \Gamma(-2\ep) s^{-1+\ep}(-u)^{-1-\ep} i}{(s+u)^{\ep} \Gamma(1-\ep) \Gamma(1-2\ep)}
  ~{}_3F_2\left(1,-2\ep,-2\ep;1-2\ep,1-\ep;1+\frac{u}{s}\right).
\end{align}

Finally, we can map the remaining integral onto a linear combination of standard Euler integrals via connection formula (\ref{eq:2F1-connection}),
\begin{align}
\label{eq:box-cut-res4}
  \Graph{.3}{cutoneloopbox} &= \frac{2\pi \Gamma(-\ep) s^{-\ep} i}{s u \Gamma(-2\ep)} \int_0^1 {\rm d} t~ (1-t)^{-1-\ep} \left(1-\left(1+\frac{s}{u}\right)t\right)^{\ep}
  \nonumber \\
  &+\frac{2\pi s^{-1+\ep}(-u)^{-1-\ep} i}{(s+u)^{\ep} \Gamma(1-\ep)}\int_0^1 {\rm d} t~ (1-t)^{-1-2\ep} {}_2F_1\left(1,-2\ep;1-\ep;(1-t)\left(1+\frac{u}{s}\right)\right)
  \nonumber \\
  &- \frac{2\pi \Gamma(-2\ep) s^{-1+\ep}(-u)^{-1-\ep} i}{(s+u)^{\ep} \Gamma(1-\ep) \Gamma(1-2\ep)}
  ~{}_3F_2\left(1,-2\ep,-2\ep;1-2\ep,1-\ep;1+\frac{u}{s}\right)
  \nonumber \\
  &= \frac{2 i \sin(\pi \ep) s^{-\ep}\Gamma^2(-\ep)\Gamma(\ep)}{s u \Gamma(-2\ep)} \,{}_2F_1\left(1,-\ep;1-\ep;1+\frac{s}{u}\right)
  \nonumber \\
  &+\frac{2\pi s^{-1+\ep}(-u)^{-1-\ep} i}{(s+u)^{\ep} \Gamma(1-\ep)}\int_0^1 {\rm d} r~ r^{-1-2\ep} {}_2F_1\left(1,-2\ep;1-\ep;r\left(1+\frac{u}{s}\right)\right)
  \nonumber \\
  &- \frac{2\pi \Gamma(-2\ep) s^{-1+\ep}(-u)^{-1-\ep} i}{(s+u)^{\ep} \Gamma(1-\ep) \Gamma(1-2\ep)}
  ~{}_3F_2\left(1,-2\ep,-2\ep;1-2\ep,1-\ep;1+\frac{u}{s}\right),
\end{align}
where we have obtained the second equality by evaluating the integral on the first line of Eq. (\ref{eq:box-cut-res4}) with the help of (\ref{eq:2F1-intdef}). Applying Eq. (\ref{eq:3F2-intdef}) to the final integral remaining on the right-hand side of (\ref{eq:box-cut-res4}), we see that the cut Feynman integral evaluates to
\begin{align}
\label{eq:box-cut-finalres}
  \Graph{.3}{cutoneloopbox} &= \frac{2 i \sin(\pi \ep) s^{-\ep}\Gamma^2(-\ep)\Gamma(\ep)}{s u \Gamma(-2\ep)} \,{}_2F_1\left(1,-\ep;1-\ep;1+\frac{s}{u}\right).
\end{align}
Note that the final integral on the right-hand side of (\ref{eq:box-cut-res4}) exactly cancels the ${}_3F_2$ term, thereby removing all dependence on the generalized hypergeometric series from the result. One can check using (\ref{eq:box-cut-finalres}) and the physical region ($s > 0$ and $-s < u < 0$) evaluation of (\ref{eq:box-def}) given in Eq. (\ref{eq:one-loop-box-virt}) that
\begin{align}
{\rm Disc}_{s}\left(\Graph{.3}{virtoneloopbox}\right)  = -\left(\Graph{.3}{cutoneloopbox}\right).
\end{align}

\subsection{The one-external-mass six-line two-loop double triangle}
\label{subsec:2formfac}
Our final example will be the $s$-channel cut of the one-external-mass six-line two-loop double triangle,
\begin{align}
\label{eq:double-triangle-def}
  \Graph{.3}{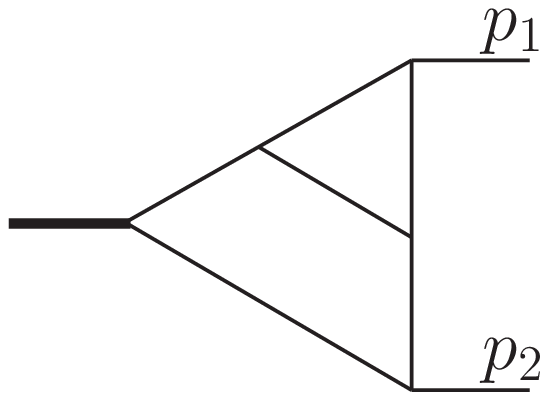} = \int \frac{d^d k_1}{i \pi^{d/2}}\int \frac{d^d k_2}{i \pi^{d/2}}~\frac{1}{ (p_1+k_1)^2 (p_1+k_2)^2 (k_1-k_2)^2 k_2^2 k_1^2 (p_2-k_1)^2}.
\end{align}
The $s$-channel cut of this integral was chosen because three cut Feynman integrals contribute to it and one of these integrals has an integration domain which is non-trivial to determine. This example will show the reader what Baikov-Lee computations look like beyond one loop, where irreducible scalar products and real-virtual contributions come into play for the first time. As before, we begin with Eq. (\ref{eq:leerepcut}). This time, however, we must enumerate the distinguishable cut Feynman integrals which contribute. Let us consider the triple cut first and the double cut second. The conjugate of the double-cut contribution can be obtained from the double cut without any additional calculation.\footnote{By virtue of the $+i 0$ in Eq. (\ref{eq:gen-virt-int-M}), the double cut contribution and its conjugate are distinguishable.}

After performing some trivial manipulations, we obtain a cut Baikov-Lee representation of the form
\begin{align}
\label{eq:triple-cut-def}
  &\Graph{.3}{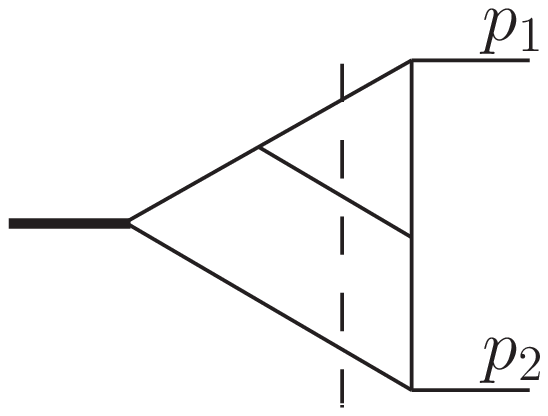} = -\frac{2^{2-4\ep} s^{-1+2\ep} i}{\Gamma (1-2\ep)} \underset{\bar{\mathcal{D}}_{s{\rm -cut}}^{(1)}}{\int \int \int \int} {\rm d}s_1{\rm d}s_2{\rm d}s_3{\rm d}s_4~\underset{{\{\bar{s}_1,\bar{s}_2,\bar{s}_3\}}}{\mathbf{Res}}\left\{\vphantom{\frac{\Big(-s^2/4 ( \bar{s}_2^2 - \bar{s}_1 s_2 ) + s ( \bar{s}_1 \bar{s}_3 s_1 - \bar{s}_2 (\bar{s}_3 s_3+s_1 s_4) + s_2 s_3 s_4 ) - (\bar{s}_3 s_3 - s_1 s_4)^2\Big)^{-1/2-\ep}}{\left(s_2+2 s_1\right)\left(\bar{s}_1+2 s_3\right)\left(\bar{s}_2-s_2/2-\bar{s}_1/2\right)\bar{s}_1 s_2 \left(\bar{s}_3-s_2/2\right)}}
  \right. \\
  &\left. \frac{\Big(-s^2 ( \bar{s}_2^2 - \bar{s}_1 s_2 )/4 + s ( \bar{s}_1 \bar{s}_3 s_1 - \bar{s}_2 (\bar{s}_3 s_3+s_1 s_4) + s_2 s_3 s_4 ) - (\bar{s}_3 s_3 - s_1 s_4)^2\Big)^{-1/2-\ep}}{\left(s_2+2 s_1\right)\left(\bar{s}_1+2 s_3\right)\left(\bar{s}_2-s_2/2-\bar{s}_1/2\right)\bar{s}_1 s_2 \left(\bar{s}_3-s_2/2\right)}\right\} \nonumber
\end{align}
for the triple-cut contribution. In Eq. (\ref{eq:triple-cut-def}), we have made the definitions $s_1 = k_1 \cdot p_1$, $s_2 = k_1^2$, $s_3 = k_2 \cdot p_1$, $s_4 = k_2 \cdot p_2$, $\bar{s}_1 = k_2^2$, $\bar{s}_2 = k_1 \cdot k_2$, $\bar{s}_3 = k_1 \cdot p_2$, and $s = (p_1 + p_2)^2$. Carrying out the residue computations, we find
\begin{align}
\label{eq:triple-cut-res1}
  &\Graph{.3}{cuttwolooptriangle} = \frac{2^{1-4\ep} s^{-1+2\ep} i}{\Gamma (1-2\ep)} \underset{\bar{\mathcal{D}}_{s{\rm -cut}}^{(1)}}{\int \int \int \int} \frac{{\rm d}s_1{\rm d}s_2{\rm d}s_3{\rm d}s_4}{s_2 s_3 \left(s_2+2 s_1\right)}~\underset{{\{\bar{s}_2,\bar{s}_3\}}}{\mathbf{Res}}\left\{\vphantom{\frac{\Big(-s^2 ( \bar{s}_2^2 + 2 s_2 s_3 )/4 + s ( s_2 s_3 s_4 - \bar{s}_2 (\bar{s}_3 s_3+s_1 s_4) - 2 \bar{s}_3 s_1 s_3 ) - (\bar{s}_3 s_3 - s_1 s_4)^2\Big)^{-1/2-\ep}}{\left(\bar{s}_2-s_2/2+s_3\right) \left(\bar{s}_3-s_2/2\right)}}
  \right. \\
  &\left. \frac{\Big(-s^2 ( \bar{s}_2^2 + 2 s_2 s_3 )/4 + s ( s_2 s_3 s_4 - \bar{s}_2 (\bar{s}_3 s_3+s_1 s_4) - 2 \bar{s}_3 s_1 s_3 ) - (\bar{s}_3 s_3 - s_1 s_4)^2\Big)^{-1/2-\ep}}{\left(\bar{s}_2-s_2/2+s_3\right) \left(\bar{s}_3-s_2/2\right)}\right\}
  \nonumber \\
  &\qquad \qquad \quad \,= \frac{2^{1-4\ep} s^{-1+2\ep} i}{\Gamma (1-2\ep)} \underset{\bar{\mathcal{D}}_{s{\rm -cut}}^{(1)}}{\int \int \int \int} \frac{{\rm d}s_1{\rm d}s_2{\rm d}s_3{\rm d}s_4}{s_2 s_3 \left(s_2+2 s_1\right)}~\underset{{\{\bar{s}_3\}}}{\mathbf{Res}}\left\{\frac{1}{\left(\bar{s}_3-s_2/2\right)} \Big(- (\bar{s}_3 s_3 - s_1 s_4)^2
  \right. \nonumber \\
  &\left.\qquad \qquad + s ( s_2 s_3 s_4 - (s_2 - 2 s_3) (\bar{s}_3 s_3+s_1 s_4)/2 - 2 \bar{s}_3 s_1 s_3 ) -s^2 ( s_2 + 2 s_3 )^2/16 \Big)^{-1/2-\ep} \vphantom{\frac{1}{\left(\bar{s}_3-s_2/2\right)}}\right\}
  \nonumber \\
  &\qquad \qquad \quad \,= \frac{2^{1-4\ep} s^{-1+2\ep} i}{\Gamma (1-2\ep)} \underset{\bar{\mathcal{D}}_{s{\rm -cut}}^{(1)}}{\int \int \int \int} \frac{{\rm d}s_1{\rm d}s_2{\rm d}s_3{\rm d}s_4}{s_2 s_3 \left(s_2+2 s_1\right)}~\Big(- (s_2 s_3 - 2 s_1 s_4)^2/4
  \nonumber \\
  &\qquad \qquad + s ( s_2 s_3 s_4 - (s_2 - 2 s_3) (s_2 s_3 + 2 s_1 s_4)/4 - s_1 s_2 s_3 ) - s^2 ( s_2 + 2 s_3 )^2/16 \Big)^{-1/2-\ep}.
  \nonumber
\end{align}

We begin our treatment of the unbarred integration variables with $s_4$. Following the discussion of Section \ref{subsec:B-L-continuation}, we obtain our integration domain by studying the zeros of polynomial structures inside the integrand. Due to the fact that the polynomial
\begin{align}
\label{eq:poly}
- (s_2 s_3 - 2 s_1 s_4)^2/4+ s ( s_2 s_3 s_4 - (s_2 - 2 s_3) (s_2 s_3 + 2 s_1 s_4)/4 - s_1 s_2 s_3 ) - s^2 ( s_2 + 2 s_3 )^2/16
\nonumber
\end{align}
is raised to the power $-1/2-\ep$, the integrand obtained above in (\ref{eq:triple-cut-res1}) has the branch points
\begin{align}
s^\pm_4 = \frac{(s_1 s_2 + s (s_1 + s_2) ) s_3 - s_1 s_2 s/2 \pm \sqrt{s} \sqrt{s_2 (2 s_1 + s_2) s_3 (s_3 - s_1) (2 s_1 + s)}}{2 s_1^2}
\end{align}
in $s_4$. As in the previous example, we can actually deduce the limits of integration on the remaining variables by simply looking at the branch cut structure of $s_4^\pm$.\footnote{It is worth pointing out that our logic here is similar to that used to formulate the compatibility graphs method for purely-virtual Feynman integrals~\cite{Brown:2009ta}. Namely, we do not expect new singularity structures to arise beyond those which are already encoded in the analytical structure of the initial integrand.} First, the polynomial structures $s_2(2 s_1 + s_2)$ and $s_3(s_3 - s_1)$ under the radical tell us that the limits of integration for $s_2$ are either $[0,-2 s_1]$ or $[-2 s_1,0]$ and that the limits of integration for $s_3$ are either $[0,s_1]$ or $[s_1,0]$. The only question that remains is whether $s_1$, what we shall consider to be the final variable of integration, is positive or negative.\footnote{All possible limits of integration for $s_2$ and $s_3$ force one of the $s_1$ integration limits to be $0$.} In the physical region, the polynomial $2 s_1 + s$ under the radical implies that the variable $s_1$ is a negative number which runs between $-s/2$ and $0$. Finally, we conclude that the $s_2$ integration runs between $0$ and $-2 s_1$ and that the $s_3$ integration runs between $s_1$ and $0$. The upshot is that
\begin{align}
\label{eq:triple-cut-limits}
  &\Graph{.3}{cuttwolooptriangle} = \frac{2^{1-4\ep} s^{-1+2\ep} i}{\Gamma (1-2\ep)} \int_{-s/2}^0 {\rm d}s_1\int_0^{-2 s_1} {\rm d}s_2\int_{s_1}^0 {\rm d}s_3\int_{s_4^-}^{s_4^+}{\rm d}s_4 \,\times
  \nonumber \\
  &\qquad \qquad \times \frac{1}{s_2 s_3 \left(s_2+2 s_1\right)} \Big(- (s_2 s_3 - 2 s_1 s_4)^2/4 - s^2 ( s_2 + 2 s_3 )^2/16 
  \nonumber \\
  &\qquad \qquad + s ( s_2 s_3 s_4 - (s_2 - 2 s_3) (s_2 s_3 + 2 s_1 s_4)/4 - s_1 s_2 s_3 )  \Big)^{-1/2-\ep}.
\end{align}

In this case, all of the remaining integrations are elementary. The $s_4$ integration is closely analogous to the integration with respect to $q_1\cdot q_2$ carried out in Appendix \ref{app:Baikov} and the other integrations may be carried out with the help of a computer algebra system such as {\tt Mathematica}. We find
\begin{align}
\label{eq:triple-cut-res-final}
  \Graph{.3}{cuttwolooptriangle} &= \frac{2 \pi s^{-1+\ep} i}{\Gamma^2(1-\ep)} \int_{-s/2}^0 {\rm d}s_1 (-s_1)^{-1+2\ep}(2 s_1 + s)^{-\ep}\int_0^{-2 s_1} {\rm d}s_2 s_2^{-1-\ep} \, \times
  \nonumber \\
  & \times \int_{s_1}^0 {\rm d}s_3 \Big(s_3-s_1\Big)^{-\ep}\Big(s_3 (2 s_1 + s_2)\Big)^{-1-\ep}
  \nonumber \\
  &= \frac{2 \pi s^{-1+\ep} i}{\ep \Gamma(1-2\ep)} \int_{-s/2}^0 \frac{{\rm d}s_1}{s_1} (2 s_1 + s)^{-\ep}\int_0^{-2 s_1} {\rm d}s_2\,s_2^{-1-\ep} (-2 s_1 - s_2)^{-1-\ep}
  \nonumber \\
  &= \frac{2^{-1-2\ep} \pi \Gamma^2(-\ep)s^{-1+\ep} i}{\ep^2 \Gamma^2(-2\ep)} \int_{-s/2}^0 {\rm d}s_1 (-s_1)^{-2-2\ep}(2 s_1 + s)^{-\ep}
  \nonumber \\
  &= -\frac{2 i \sin(2 \pi \ep) s^{-2-2 \ep} \Gamma(-1-2\ep)\Gamma(1+2\ep)\Gamma^3(-\ep)}{\Gamma(1-2\ep)\Gamma(-3\ep)}
\end{align}
for the triple-cut contribution.

Our next task is to calculate the double-cut contribution. In fact, we can write down the answer immediately by recycling calculations that we have already carried out. We require only the result obtained in Section \ref{subsec:triangle-gendots} for the $s$-channel cut of the one-external-mass one-loop triangle integral together with the well-known result for its purely-virtual counterpart. The key observation is that, on the support of the double cut, the virtual part of the cut one-external-mass six-line two-loop double triangle is precisely a purely-virtual one-loop triangle with external mass $k_1^2$. From Eq. (\ref{eq:one-loop-triangle-virt}), we see that\footnote{Here, the $+i 0$ prescription
for the propagator plays a crucial role.}
\begin{align}
\label{eq:double-cut-virtual}
  \Graph{.3}{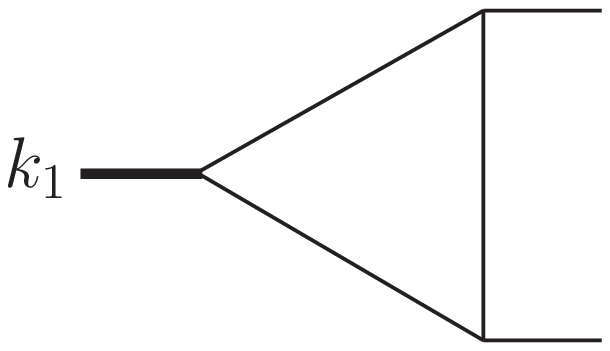} = \frac{e^{i \pi \ep} \Gamma(1+\ep) \Gamma^2(-\ep)}{\Gamma(1-2\ep)\left(k_1^2\right)^{1+\ep}}.
\end{align}
It therefore follows that we can treat the integration over $k_1$ using Eq. (\ref{eq:one-loop-triangle-cut}) with propagator exponents $\nu_1 = 1$, $\nu_2 = 1$, and $\nu_3 = 2 + \ep$. The desired result is
\begin{align}
\label{eq:double-cut-virtual-res}
  \Graph{.3}{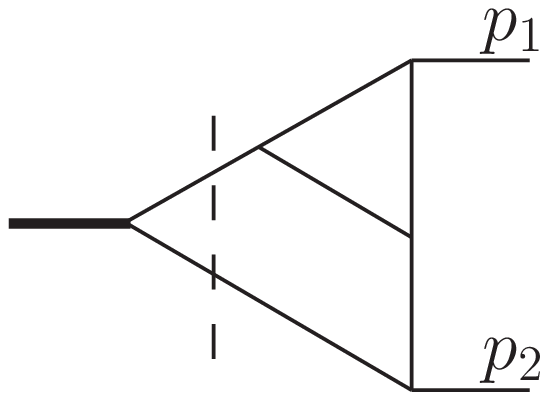} &= \frac{e^{i \pi \ep} \Gamma(1+\ep) \Gamma^2(-\ep)}{\Gamma(1-2\ep)} \left(\Graph{.3}{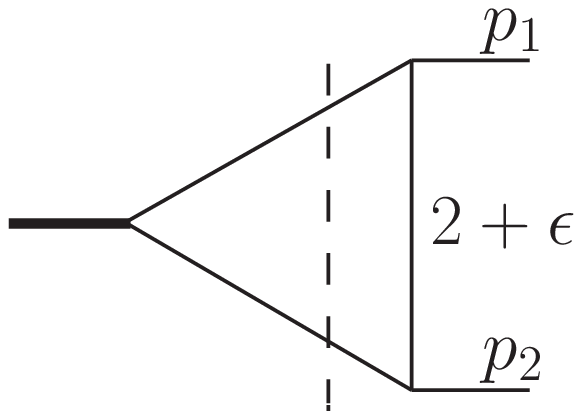}\right)
  \nonumber \\
  &= -\frac{2 i e^{i \pi \ep} \sin(\pi \ep) s^{-2-2\ep} \Gamma^2(-1-2\ep)\Gamma(1+\ep)\Gamma(2+2\ep)\Gamma^2(-\ep)}{\Gamma(-3\ep)\Gamma(1-2\ep)}.
\end{align}

The final cut Feynman integral of interest is the conjugate double-cut contribution. The result may be simply obtained by taking the complex conjugate of the virtual part of the double-cut contribution. We have
\begin{align}
\label{eq:double-cut-virtual-conjres}
  \Graph{.3}{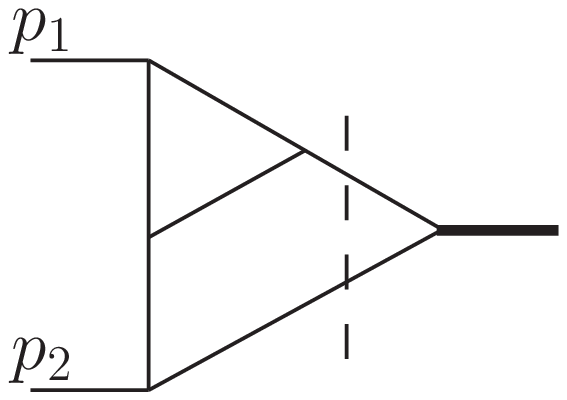} &= \frac{e^{-i \pi \ep} \Gamma(1+\ep) \Gamma^2(-\ep)}{\Gamma(1-2\ep)} \left(\Graph{.3}{cutonelooptrianglespec}\right)
  \nonumber \\
  &= -\frac{2 i e^{-i \pi \ep} \sin(\pi \ep) s^{-2-2\ep} \Gamma^2(-1-2\ep)\Gamma(1+\ep)\Gamma(2+2\ep)\Gamma^2(-\ep)}{\Gamma(-3\ep)\Gamma(1-2\ep)}.
\end{align}
Finally, the sum of the three contributions to the $s$-channel cut, Eqs. (\ref{eq:triple-cut-res-final}), (\ref{eq:double-cut-virtual-res}), and (\ref{eq:double-cut-virtual-conjres}), may be rewritten as
\begin{align}
\label{eq:two-loop-triangle-cut}
  &\Graph{.3}{cuttwolooptriangle}+\Graph{.3}{cuttwolooptriangleRV}+\Graph{.3}{cuttwolooptriangleRVflip} = 
  \\
  &\qquad \qquad -\frac{2 i \sin(2 \pi \ep) s^{-2-2 \ep} \Gamma(-1-2\ep)\Gamma(1+2\ep)\Gamma^2(-\ep)}{\Gamma(1-2\ep)\Gamma(-3\ep)}\Big(\Gamma(-\ep) - \Gamma(1+\ep)\Gamma(-2\ep)\Big).
  \nonumber
\end{align}
Using (\ref{eq:two-loop-triangle-cut}) and the $s > 0$ evaluation of (\ref{eq:double-triangle-def}) given in Eq. (\ref{eq:two-loop-triangle-virt}), the reader can check that
\begin{align}
{\rm Disc}_{s}\left(\Graph{.3}{virttwolooptriangle}\right)  = -\left(\Graph{.3}{cuttwolooptriangle}+\Graph{.3}{cuttwolooptriangleRV}+\Graph{.3}{cuttwolooptriangleRVflip}\right).
\end{align}


\section{Maximally-cut Feynman integrals and differential equations}
\label{sec:maxcut}

As mentioned in the introduction, it was observed in~\cite{Primo:2016ebd} that maximally-cut Feynman integrals~\cite{Bern:2004ky,Britto:2004nc} in the $\ep \to 0$ limit solve the homogeneous parts of the differential equations satisfied by their uncut counterparts. In non-polylogarithmic cases, it was furthermore shown explicitly that maximally-cut Feynman integrals in the $\ep \to 0$ limit may often be computed by direct integration. However, complete solution sets of their homogeneous differential equations were obtained indirectly by using a convenient mathematical property of the complete elliptic integral of the first kind. Subsequently, it was shown in~\cite{Frellesvig:2017aai} that Baikov's method may be applied to maximally-cut Feynman integrals and offers conceptual advantages relative to the traditional approach to maximally-cut Feynman integrals. However only one solution of their higher-order homogeneous differential equations was provided. 

We explain in this section how one can directly obtain the necessary, complete solution sets for a class of interesting non-polylogarithmic Feynman integrals. As we shall see, the key idea is to again employ our cut Baikov-Lee representation, Eq. (\ref{eq:leerepcut}), but to allow for more general integration domains than have so far been considered. For all of the classical unitarity cuts studied in Section \ref{sec:stdcuts}, a unique solution was obtained via a generalized phase-space volume computation. However, the criteria that we employed to determine the integration domain must now be generalized further to allow for multiple solutions, due to the fact that the differential equations satisfied by non-polylogarithmic Feynman integrals are of order greater than one. Although the primary focus of this section will be on maximally-cut Feynman integrals which evaluate to generic complete elliptic integrals, the ideas discussed here apply to maximally-cut Feynman integrals which evaluate to complete hyperelliptic integrals as well~\cite{Carlson:2003}.

Let us consider the following family of elliptic curves,
\begin{equation}
\label{eq:sqrt}
y(x,t)=\sqrt{\Big(x-a_1(t)\Big)\Big(x-a_2(t)\Big)\Big(x-a_3(t)\Big)\Big(x-a_4(t)\Big)}\,,
\end{equation}
where $t$ is a parameter and $\{a_i(t)\}$ is the set of branch points of $y(x,t)$, with $a_i(t)\neq a_j(t)$ for $i\neq j$. Now, for some Feynman integral, $I$, let us suppose that we are able to obtain a one-fold integral representation for its maximal cut, $\bar{I}$, of the form
\begin{equation}
\label{eq:elliptic-integral}
\bar{I}=\underset{\bar{\mathcal{D}}_{\{\bar{s}_i\}}}{\int} {\rm d}x \, R(x,y(x,t),t),
\end{equation}
where $R(x,y(x,t),t)$ is a rational function of its arguments. If $R(x,y(x,t),t)$ has no poles in $x$, it is then the case that a complete set of solutions to the homogeneous part of the differential equations satisfied by $I$ may be obtained by considering
\begin{equation}
\label{eq:integral-branching}
\bar{I}(a_i(t),a_j(t))=\int_{a_i(t)}^{a_j(t)} {\rm d}x \, R(x,y(x,t),t)
\end{equation}
for $i\neq j$. If poles are present, one must also consider closed contour integrals around each pole, 
\begin{equation}
\label{eq:integral-poles}
\bar{I}(\gamma_k)=\oint_{\gamma_k} {\rm d}x \, R(x,y(x,t),t),
\end{equation}
to find all possible solutions. In (\ref{eq:integral-poles}), $\gamma_k$ denotes a closed contour which encircles the $k$-th pole of $R(x,y(x,t),t)$, but no other pole or branch point of the integrand.  Note that, quite generically, the set of solutions obtained in this manner will actually be overcomplete.

By using properties of elliptic curves we can argue that (\ref{eq:integral-branching}) and  (\ref{eq:integral-poles}) represent a complete set of solutions. Integrals $\bar{I}(a_i(t),a_j(t))$ for $i\neq j$ and $\bar{I}(\gamma_k)$ are periods of the elliptic curve $y(x,t)$~\cite{Kontsevich:2001} and the given homogeneous differential equation for $I$ with respect to $t$ is nothing but the associated Picard-Fuchs equation.\footnote{It has been clear for a long time that Picard-Fuchs equations play a very important role in the theory of Feynman integrals (see {\it e.g.}~\cite{MullerStach:2011ru,MullerStach:2012mp}).} By construction, these differential equations are the same for every period, and a complete set of periods provides a complete set of solutions to the Picard-Fuchs equation (for a comprehensive review of the subject of periods see~\cite{Kontsevich:2001} and the references therein).

The prescription described above represents a substantial generalization of that described in Section \ref{subsec:B-L-continuation}. When considering a cut Feynman integral associated with the unitarity cut in some physical kinematic channel, one naturally expects to obtain a result which is real-valued up to an overall phase; in this context, however, no such constraint applies and the integration domain is no longer uniquely determined. Indeed, Eq.~(\ref{eq:integral-branching}) implies that, in the elliptic case, we can integrate the maximal cut of $I$ over six distinct domains and this obviously leads to some solutions which possess both real and imaginary parts. If the integrand has poles one must also include solutions of the form (\ref{eq:integral-poles}). This would be relevant, for example, when considering a complete elliptic integral of the third kind. 

In the absence of poles, complete elliptic integrals admit a simple description as closed contour integrals which wrap the torus~\cite{Carlson:2003}. As claimed above, it is clear from this point of view that our solutions cannot form a linearly independent set. To see this, recall that the fundamental group of the torus, $\mathbb{Z}\times \mathbb{Z}$, is isomorphic to the first homology group (Hurewicz's theorem~\cite{Hatcher:2002}). This means that, in the absence of poles, the torus admits just two independent cycles for us to integrate along. We can therefore conclude that four of the six functions generated by applying the prescription given in (\ref{eq:integral-branching}) above are actually spurious and may be disposed of.
In general, one must also check whether contour integrals of the form (\ref{eq:integral-poles}) around different poles of the integrand yield linearly dependent results. 

The above discussion generalizes and systematizes the analysis of, {\it e.g.}, \cite{Adams:2013kgc} to generic complete elliptic integrals of the form~(\ref{eq:elliptic-integral}). Moreover, it is possible to generalize it to curves of higher genus, {\it i.e.} when the square of (\ref{eq:sqrt}) is a polynomial of degree greater than four, by considering the set of periods over the relevant higher-genus Riemann surface. The application of these techniques to curves of genus greater than one goes beyond the scope of the present paper but will likely play a role in future calculations. In the following, we consider illustrative examples taken from the virtual corrections to Higgs + jet with exact top mass dependence.

\subsection{A Higgs + jet non-polylogarithmic three-point function}
\label{sec:maxcut-3pt}
As a first example, we consider the maximal cut of the two-loop crossed form factor,
\begin{equation}
\label{eq:crossed-triangle}
  \Graph{.3}{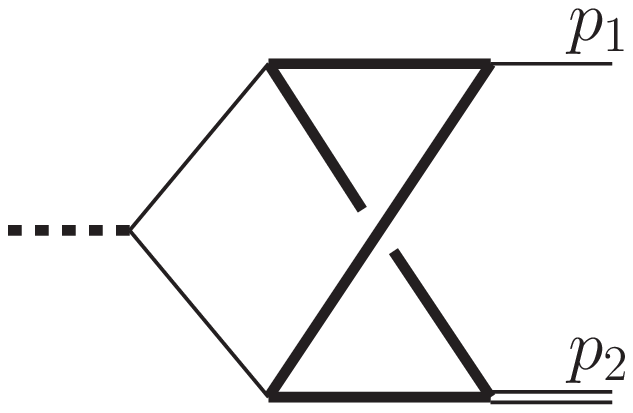}=\int \frac{{\rm d}^d k_1}{i \pi ^{d/2}}\frac{{\rm d}^d k_2}{i \pi ^{d/2}} \prod_{i=1}^{6} D_i^{-1},
\end{equation}
where we have made the definitions,
\begin{align}
  D_1 & =(k_2+p_1)^2-m^2 \qquad & D_2 & = (k_1-k_2-p_1-p_2)^2 \qquad & D_3 & = k_2^2-m^2 
  \nonumber \\
  D_4 & =(k_1-p_2)^2-m^2 \qquad & D_5 & = (k_1-k_2)^2 \qquad & D_6 &  = k_1^2-m^2.
\end{align}
The evaluation of this Feynman integral is relevant to the calculation of the non-planar part of the two-loop virtual corrections to Higgs + jet with exact top mass dependence. Its maximal cut was considered in \cite{Primo:2016ebd}, where it was evaluated with a traditional cut parametrization in an effort to obtain a solution to the homogeneous part of the associated system of differential equations.
Using Eq.~(\ref{eq:leerepcut}), we arrive at the following one-fold integral representation of the maximally-cut Feynman integral,
\begin{align}
  \Graph{.3}{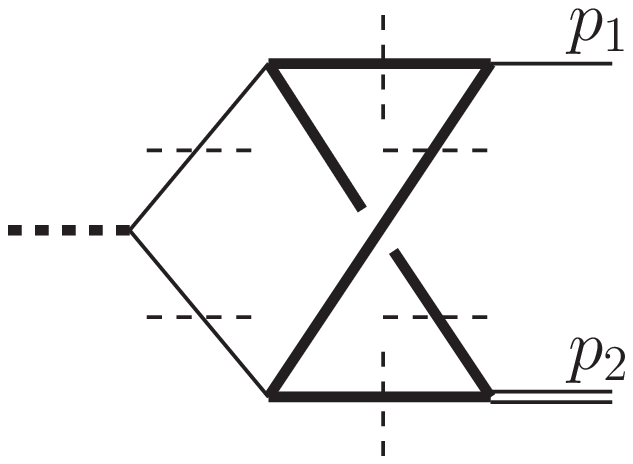} &= \frac{2^{4-2\epsilon} \pi^3}{(s-p_2^2)^{1-2\ep}\Gamma \left(1-2\epsilon\right)} \times
  \nonumber \\ 
  & \times \underset{\bar{\mathcal{D}}_{\{\bar{s}_i\}}}{\int} {\rm d}s_1 \left(s_1 \left(s-p_2^2+2 s_1\right) \left(2 m^2 s-s_1 \left(s-p_2^2+2 s_1\right)\right)\right)^{-1/2-\epsilon},
\end{align}
where we have made the definitions $s_1 =  k_1\cdot p_1$ and $s = (p_1 + p_2)^2$, and we have chosen to work in the physical region above threshold where $s > p_2^2 > 0$ and $m^2 > 0$.

We now take the first step towards finding a complete set of solutions to the homogeneous differential equation for the uncut integral by computing the maximal cut in $d=4$. We have,
\begin{align}
\label{eq:maximal-cut-crossed-triangle}
  &\underset{\ep \to 0}{\rm lim}\left\{\Graph{.3}{crossedtrianglemax}\right\} = \frac{16 \pi^3}{s-p_2^2} \underset{\bar{\mathcal{D}}_{\{\bar{s}_i\}}}{\int} \frac{{\rm d}s_1}{y\left(s_1,m^2,p_2^2,s\right)}\,,
\end{align}
where
\begin{align}
\label{eq:elliptic-curve3pt}
  y\left(s_1,m^2,p_2^2,s\right) = \sqrt{s_1 \left(s-p_2^2+2 s_1\right) \left(2 m^2 s-s_1 \left(s-p_2^2+2 s_1\right)\right)}\,.
\end{align}

Following our general prescription, concrete results are obtained by integrating between branch points of the integrand. In other words, we obtain six possible solutions by picking distinct pairs of elements from the set of branch points of the integrand,
\begin{equation}
\label{eq:branch-points}
  \left\{\frac{p_2^2-s-\rho}{4},\,\frac{p_2^2-s}{2},\,0,\,\frac{p_2^2-s+\rho}{4}\right\},
\end{equation}
where we have introduced the convenient shorthand
\begin{equation}
  \rho=\sqrt{16 m^2 s+\left(s-p_2^2\right)^2}
\end{equation}
in (\ref{eq:branch-points}). Note that, in the kinematic region that we are working in, the elements of (\ref{eq:branch-points}) are real-valued and ordered from smallest to largest.

As discussed above, the solutions to our homogeneous differential equations will be complete elliptic integrals and these may be thought of as periods of the torus. The torus admits two linearly independent cycles and we therefore expect to find just two linearly independent periods. As we shall see, it is convenient to take
\begin{align}
\label{eq:f1}
  f_1 &\equiv \frac{16 \pi^3}{s-p_2^2} ~~~~\int_{(p_2^2-s)/2}^0 \frac{{\rm d}s_1}{y\left(s_1,m^2,p_2^2,s\right)}
  \\
\label{eq:f2}
  f_2 &\equiv \frac{16 \pi^3}{s-p_2^2} \int_0^{(p_2^2-s+\rho)/4} \frac{{\rm d}s_1}{y\left(s_1,m^2,p_2^2,s\right)}
\end{align}
to be our independent basis elements. 

That $f_1$ and $f_2$ are actually independent periods may be seen by writing them in a standard form. For $f_1$, this is easily achieved by making the appropriate analytical continuation of Eq. (\ref{eq:elliptic-curve3pt}) and then changing variables according to~\cite{Primo:2016ebd}
\begin{align}
s_1 = \frac{\frac{p_2^2-s}{2} \,t^2}{1-\frac{2 \left(s-p_2^2\right)}{s-p_2^2-\rho}\,(1-t^2)}. \nonumber
\end{align}
The result is
\begin{align}
\label{eq:f1-res}
  f_1 =  -\frac{64 \pi^3 i}{\left(s-p_2^2\right)\left(s-p_2^2+\rho\right)}~K\left(\frac{4 \rho \left(s-p_2^2\right)}{\left(s-p_2^2+\rho\right)^2}\right)
\end{align}
in the physical kinematic region of interest.\footnote{In Eqs. (\ref{eq:f1-res}) and (\ref{eq:f2-res}), $K(z)$ is the complete elliptic integral of the first kind,
\begin{align}
  K(z) = \int_0^1 \frac{{\rm d} t}{\sqrt{(1-t^2)(1-z \, t^2)}}\,.\nonumber
\end{align}}
Similar considerations lead to
\begin{align}
\label{eq:f2-res}
  f_2 = \frac{32 \pi^3}{\left(s-p_2^2\right) \sqrt{\rho \left(s-p_2^2\right)}}~ K\left(-\frac{\left(s-p_2^2-\rho\right)^2}{4 \rho \left(s-p_2^2\right)}\right),
\end{align}
again in the kinematic region of interest. The other possible solutions may be written as linear combinations of $f_1$ and $f_2$ and it is clear that some of them will have both real and imaginary parts. For example, we have from (\ref{eq:branch-points}) and the explicit expressions for $f_1$ and $f_2$ that
\begin{align}
\frac{16 \pi^3}{s-p_2^2} \int_{(p_2^2-s)/2}^{(p_2^2-s+\rho)/4} \frac{{\rm d}s_1}{y\left(s_1,m^2,p_2^2,s\right)} = f_1 + f_2,
\end{align}
where $f_1$ is purely imaginary and $f_2$ is purely real in the region where $s > p_2^2 > 0$ and $m^2 > 0$.

Finally, let us point out that we have explicitly checked that the maximal cut calculated in this section solves the associated homogeneous differential equation. A more non-trivial Higgs + jet four-point function is discussed in the next section as a further application of our ideas. The following calculation demonstrates the utility of integrating out cut loops one at a time. As pointed out in~\cite{Frellesvig:2017aai}, working in this way often allows one to write down a more compact integrand than in the straightforward all-at-once approach to the construction of a cut Baikov-Lee representation adopted so far in this paper.

\subsection{A Higgs + jet non-polylogarithmic four-point function}
\label{subsec:maxcut-4pt}
We consider the following two-loop Higgs + jet integral, denoted in \cite{Bonciani:2016qxi} as $f^A_{66}$,
\begin{align}
\label{eqn:higgsjet}
  \Graph{.3}{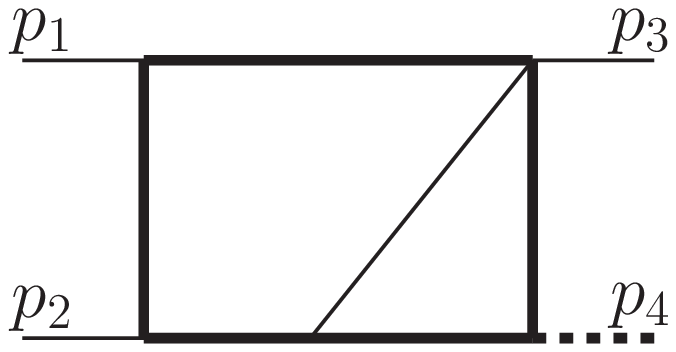} =\int \frac{{\rm d}^d k_1}{i \pi ^{d/2}}\frac{{\rm d}^d k_2}{i \pi ^{d/2}} \prod_{i=1}^{6} D_i^{-1},
\end{align}
with propagators,
\begin{align}
   D_1 & = (k_1+p_3+p_4)^2-m^2 \quad & D_2 & = (k_1+p_1)^2-m^2 \quad & D_3 & = k_1^2-m^2  \nonumber \\
   D_4 & = (k_2+k_1+p_3+p_4)^2-m^2 \quad & D_5 & = (k_2+k_1+p_3)^2-m^2 \quad & D_6 &  = k_2^2. 
\end{align}
As usual, the kinematics is
\begin{align}
  s=(p_1+p_2)^2\qquad t=(p_1-p_3)^2\qquad u=(p_1-p_4)^2\qquad p_4^2=s+t+u.
\end{align} 
The maximal cut of this Feynman integral was also considered more recently in both references~\cite{Primo:2016ebd} and~\cite{Frellesvig:2017aai}. For the purposes of our analysis in this section, it is convenient to work in the kinematic region where $s > p_4^2 > 0$, $s > 4 m^2 > 0$, and $p_4^2 - s > t$.

As we are considering a four-point function at two loops, there are nine scalar product integration variables, and it would therefore seem that we must consider a three-fold integral representation of the maximal cut. Fortunately, one can obtain a one-fold integral representation of the maximal cut by proceeding recursively loop-by-loop. First, we integrate out the one-loop triangle subintegral defined by the propagators $D_4$, $D_5$, and $D_6$,
\begin{align}
\label{eq:onelooptriangle}
\Graph{.3}{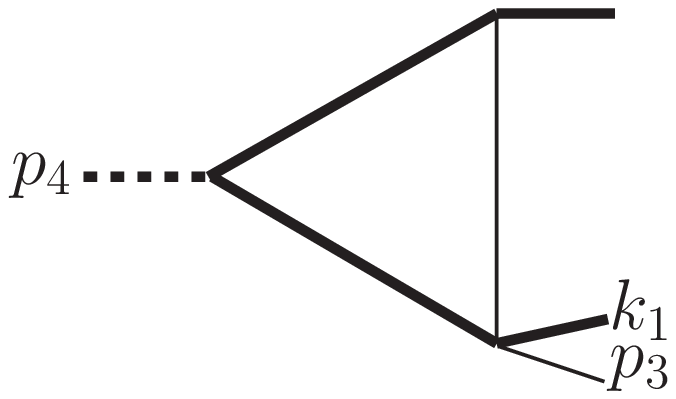}=\int \frac{{\rm d}^d k_2}{i \pi^{d/2}} \frac{1}{k_2^2\left((k_2+k_1+p_3)^2-m^2\right)\left((k_2+k_1+p_3+p_4)^2-m^2\right)},
\end{align}
by localizing $k_2^2$, $k_2 \cdot p_3$, and $k_2 \cdot p_4$. To do so, we evaluate the maximal cut of (\ref{eq:onelooptriangle}) using  Eq.~(\ref{eq:leerepcut}), our cut Baikov-Lee representation. Of course, the maximal cut of a one-loop triangle involves no non-trivial integrations and one immediately finds
\begin{align}
  &\Graph{.3}{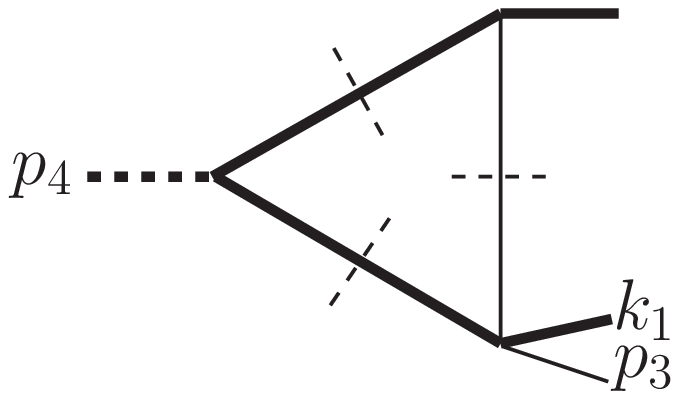} = \frac{4 \pi ^2}{\Gamma (1-\epsilon)}\left(\left(s-p_4^2-2 \,k_1 \cdot p_4\right)^2+4 \left(k_1^2+2 \,k_1 \cdot p_3\right)p_4^2\right)^{-1/2+\ep}\times
   \\
  &\times\left(-m^2 \left(s-2\,k_1\cdot p_4\right)^2-p_4^2\left(k_1^2+2\,k_1\cdot p_3+m^2\right)\left(k_1^2+2 \,k_1\cdot p_3+2\, k_1\cdot p_4-s+m^2\right)\right)^{-\ep}\nonumber
\end{align}
after carrying out the residue computations.

We now integrate out the remaining loop by localizing $k_1^2$, $k_1 \cdot p_1$, and $k_1\cdot p_4$ with the remaining cut conditions. A moment's thought reveals that Eq. (\ref{eq:leerepcut}) may still be straightforwardly applied to loop-by-loop Baikov-Lee calculations; the only difference is that the results of the previous loop integrations appear in the current integrand. In other words, if we make the definitions $s_1 = k_1 \cdot p_3$, $\bar{s}_1 = k_1^2$, $\bar{s}_2 = k_1 \cdot p_1$, and $\bar{s}_3 = k_1 \cdot p_4$, we have
\begin{align}
  &\Graph{.3}{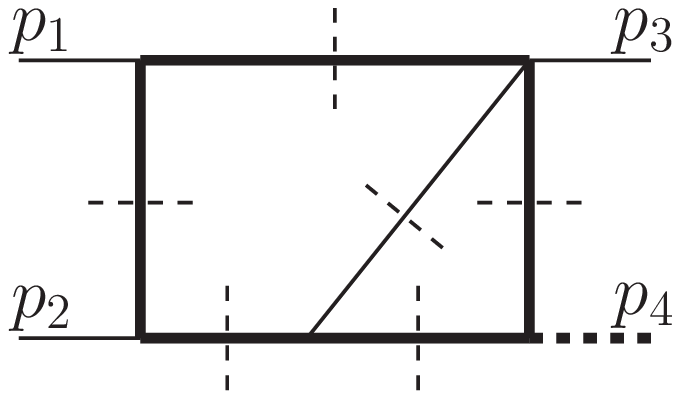}= -\frac{8 \pi ^{3/2}i}{\Gamma \left(1/2-\epsilon\right) \left[\bar{G}(p_1,p_3,p_4)\right]^{-\ep}} \underset{\bar{\mathcal{D}}_{\{\bar{s}_i\}}}{\int} {\rm d}s_1 ~\times
  \\
  &\times \underset{{\{\bar{s}_1,\bar{s}_2,\bar{s}_3\}}}{\mathbf{Res}}\left\{\vphantom{\Graph{.3}{trianglemax}}\frac{\left[\bar{G}(k_1,p_1,p_3,p_4)\right]^{-1/2-\epsilon}}{\left(\bar{s}_1+m^2\right) \left(\bar{s}_1+2 \bar{s}_2+m^2\right) \left(\bar{s}_1+2 \bar{s}_3+2 s_1-s+m^2\right)}\left(\Graph{.3}{trianglemax}\right)\right\} \nonumber
\end{align}
and can immediately write
\begin{align}
\label{eq:maximal-cut-box-triangle}
  &\underset{\ep \to 0}{\rm lim}\left\{~~\Graph{.3}{higgsjetmax}\right\} = 32 \pi^3 \underset{\bar{\mathcal{D}}_{\{\bar{s}_i\}}}{\int} \frac{{\rm d}s_1}{y\left(s_1,m^2,p_4^2,s,t\right)}\,,
\end{align}
where
\begin{align}
\label{eq:elliptic-curve4pt}
   y\left(s_1,m^2,p_4^2,s,t\right) = \sqrt{\left(p_4^2+2 s_1\right)^2 - 4 m^2 p_4^2} \sqrt{s\left(4 m^2 t \left(s+t-p_4^2\right)-s (t+2 s_1)^2\right)}
\end{align}
in the region of interest.

We now turn to the problem of finding a complete set of homogeneous solutions to the differential equations for the uncut integral, proceeding as described at the beginning of Section~\ref{sec:maxcut}.
Possible solutions are obtained by integrating between the branch points of the integrand,
\begin{align}
\label{eq:branch-points4pt}
\left\{-\frac{p_4^2}{2}-\sigma,-\frac{p_4^2}{2}+\sigma,-\frac{t}{2}-\frac{\tau}{s},-\frac{t}{2}+\frac{\tau}{s}\right\},
\end{align}
where we have set
\begin{align}
\sigma = \sqrt{m^2 p_4^2} \qquad {\rm and} \qquad \tau = \sqrt{m^2 s t \left(s+t-p_4^2\right)}\,.
\end{align}
In the kinematic region that we have chosen to work in, the elements of (\ref{eq:branch-points4pt}) are real-valued and ordered from smallest to largest. 

As guaranteed by the form of Eqs. (\ref{eq:maximal-cut-box-triangle}) and (\ref{eq:elliptic-curve4pt}), we again find just two linearly independent solutions,
\begin{align}
\label{eq:g1}
  g_1 &\equiv 32 \pi^3 \int_{-\frac{p_4^2}{2}+\sigma}^{-\frac{t}{2}-\frac{\tau}{s}}\, \frac{{\rm d}s_1}{y\left(s_1,m^2,p_4^2,s,t\right)}
\end{align}
and
\begin{align}
\label{eq:g2}
  g_2 &\equiv 32 \pi^3 \int_{-\frac{t}{2}-\frac{\tau}{s}}^{-\frac{t}{2}+\frac{\tau}{s}}\, \frac{{\rm d}s_1}{y\left(s_1,m^2,p_4^2,s,t\right)}.
\end{align}
That these two solutions are actually linearly independent follows from the explicit formulas written in terms of complete elliptic integrals of the first kind,
\begin{align}
\label{eq:g1-res}
  g_1 = -\frac{32 \pi^3 i K\left(1-\frac{16 \,\sigma \tau}{s \left(p_4^2-t\right){}^2 + 8 \,\sigma \tau - 4 \,m^2 \left(s \,p_4^2 + t \left(s+t-p_4^2\right)\right)}\right)}{\sqrt{s\left(s \left(p_4^2-t\right)^2 + 8 \,\sigma \tau - 4 \,m^2 \left(s \,p_4^2 + t \left(s+t-p_4^2\right)\right)\right)}}
\end{align}
and
\begin{align}
\label{eq:g2-res}
  g_2 = \frac{32 \pi^3 K\left(\frac{16 \,\sigma \tau}{s \left(p_4^2-t\right){}^2 + 8 \,\sigma \tau - 4 \,m^2 \left(s \,p_4^2 + t \left(s+t-p_4^2\right)\right)}\right)}{\sqrt{s\left(s \left(p_4^2-t\right)^2 + 8 \,\sigma \tau - 4 \,m^2 \left(s \,p_4^2 + t \left(s+t-p_4^2\right)\right)\right)}},
\end{align}
which may be derived by making appropriate changes of variables in Eqs. (\ref{eq:g1}) and (\ref{eq:g2}).\footnote{To derive (\ref{eq:g1-res}), one must analytically continue the second square root structure in Eq. (\ref{eq:elliptic-curve4pt}) above.}
We have explicitly checked that $g_1$ and $g_2$ satisfy the appropriate homogeneous second-order differential equations.


\section{Conclusions}
\label{sec:conc}
In this paper, we formulated and studied cut Baikov-Lee representations, both for Feynman integrals cut in a single kinematic channel and for maximally-cut Feynman integrals. For a wide class of interesting problems, our framework provides a 
convenient setup for the explicit computation of cut Feynman integrals. It makes the dependence on the Lorentz-invariant kinematic variables manifest and may be used directly or in conjunction with other methods such as sector decomposition~\cite{Binoth:2000ps,Bogner:2007cr}. Although some elements of our analysis in Section \ref{sec:genform} relied upon physically-motivated plausibility arguments and experimentation, we subsequently presented a substantial amount of evidence in Sections \ref{sec:stdcuts} and \ref{sec:maxcut} that our master formula, Eq. (\ref{eq:leerepcut}), is correct.\footnote{Although we have employed Eq. (\ref{eq:leerepcut}) throughout this paper, Eq. (\ref{eq:leerepcut-trad}) is actually on more solid ground from the theoretical point of view. The difference is that we have included all of the phases in Eq. (\ref{eq:leerepcut-trad}) which one would find by considering the relevant Feynman integrals to be Feynman graphs inside of an appropriate generalized scalar field theory. It is not clear to us why Eq. (\ref{eq:leerepcut}) works as well as it does.}

It would be very interesting in future work to consider still more non-trivial examples\footnote{When a reference evaluation is not available, it is important to check analytical results numerically. We therefore note that the recently-released program {\tt pySecDec}~\cite{Borowka:2017idc} should allow for the evaluation of a wide class of cut Feynman integrals numerically (up to some fixed order in $\ep$) with moderate user input.} such as the $s$-channel cut of the massless two-loop non-planar double box; as discussed in~\cite{vonManteuffel:2014qoa}, examples for which one cannot avoid imaginary parts on the mass shell require special care and may be instructive. In an effort to remove as many superfluous assumptions as possible from the formulation given in Section \ref{sec:genform}, it would of course also be desirable to put the theoretical foundations of the cut Baikov-Lee representation on a firmer footing. Although we have employed the familiar language of classical complex analysis throughout this work, it might be interesting to reformulate our findings in more modern language along the lines of~\cite{Abreu:2017ptx,Abreu:2017enx}. It is unclear to us, however, that such a reformulation will immediately lead to clarifications.

In fact, there exist several interesting classes of cut Feynman integrals which were not discussed in this work at all. First of all, it would be interesting to study representative sequential cuts of the type discussed in reference~\cite{Abreu:2014cla}. One should also check whether a cut-discontinuity relation of the type discussed in~\cite{Abreu:2014cla} also exists for ``crossed'' sequential cuts such as the $s$-channel $+$ $t$-channel cut of the massless one-loop box of Section \ref{subsec:1masslessbox}. Although success is less certain, it might be interesting to use the Baikov-Lee formalism developed in this work to study iterated cuts in a single channel. As a start, one could consider the Feynman integral analog of the double two-particle cuts at two loops discussed in~\cite{Bern:2000dn}. For such iterated cuts, it is not obvious that a cut-discontinuity relation exists at all, and it would therefore be interesting to take a fresh look at the problem using our Baikov-Lee machinery. Finally, it goes almost without saying that we would very much like to apply our techniques to the evaluation of the master integrals relevant to the current generation of phenomenologically-important unsolved problems in perturbative quantum field theory.


\section*{Acknowledgments}
The authors would especially like to thank Ruth Britto for many interesting discussions, support, and comments on the manuscript. We also gratefully acknowledge an illuminating discussion with Stefan M{\"u}ller-Stach and thank him for reading excerpts from our manuscript. RMS would like to thank Sven-Olaf Moch for an interesting discussion which, at least in part, inspired the author to pursue this line of research. This project has received funding from the European Research Council (ERC) under the European Union's Horizon 2020 research and innovation programme under grant agreement No 647356 (CutLoops). Our figures were generated using {\tt Jaxodraw} \cite{Binosi:2003yf}, based on {\tt AxoDraw} \cite{Vermaseren:1994je}.


\appendix


\section{Mathematical relations for hypergeometric-like functions}
\label{app:hypergeo}
In this appendix, we review some well-known facts used in this paper about hypergeometric functions and their generalizations. First of all, let us recall the usual integral representation of the hypergeometric function which provides the analytical continuation of the hypergeometric series in non-exceptional cases. For $|{\rm arg}(1-z)| < \pi$ and $\mathfrak{Re}(c) > \mathfrak{Re}(a) > 0$, we have~\cite{Lebedev:1965}
\begin{align}
    {}_2F_1(a,b;c;z)=\frac{\Gamma (c)}{\Gamma (a) \Gamma (c-a)}~ \int_0^1 {\rm d}t\, t^{a-1} (1-t)^{c-a-1} (1-t z)^{-b}.
    \label{eq:2F1-intdef}
\end{align}
In fact, for $|{\rm arg}(1-z_1)| < \pi$, $|{\rm arg}(1-z_2)| < \pi$, and $\mathfrak{Re}(c) > \mathfrak{Re}(a) > 0$, a completely analogous formula holds for the Appell $F_1$ function~\cite{Schlosser:2013}:
\begin{align}
    F_1(a;b_1,b_2;c;z_1,z_2)=\frac{\Gamma (c)}{\Gamma (a) \Gamma (c-a)}~ \int_0^1 {\rm d}t\, t^{a-1} (1-t)^{c-a-1} (1-t z_1)^{-b_1} (1-t z_2)^{-b_2}.
    \label{eq:F1-intdef}
\end{align}
We also encounter the generalized hypergeometric function ${}_3F_2$, which, for the purposes of this paper, may be defined via the integral representation
\begin{align}
    {}_3F_2(a_1,a_2,a_3;b_1,b_2;z)=\frac{\Gamma (b_2)}{\Gamma (a_3) \Gamma (b_2-a_3)}~ \int_0^1 {\rm d}t\, t^{a_3-1} (1-t)^{b_2-a_3-1} {}_2F_1(a_1,a_2;b_1;t z),
    \label{eq:3F2-intdef}
\end{align}
which is valid for $|{\rm arg}(1-z)| < \pi$ and $\mathfrak{Re}(b_2) > \mathfrak{Re}(a_3) > 0$~\cite{Lebedev:1965}.

In Lebedev~\cite{Lebedev:1965}, one also finds a very clear discussion of both linear and quadratic transformations of the hypergeometric function ${}_2F_1$. Of particular interest to us are,
\begin{align}
    {}_2F_1(a,b;c;z) &= \frac{ \Gamma (c) \Gamma (b-a)}{\Gamma (b) \Gamma (c-a)}~ (1-z)^{-a} {}_2F_1\left(a,c-b;a-b+1;\frac{1}{1-z}\right) \nonumber \\
    &\qquad+\frac{\Gamma (c) \Gamma (a-b)}{\Gamma (a) \Gamma (c-b)}~ (1-z)^{-b} {}_2F_1\left(b,c-a;b-a+1;\frac{1}{1-z}\right),
    \label{eq:2F1-connection}
\end{align}
\begin{align}
    {}_2F_1\left(a,b;2 b;\frac{4 z}{\left(1+z\right)^2}\right)=\left(1+z\right)^{2 a} \,{}_2F_1\left(a,a-b+\frac{1}{2};b+\frac{1}{2};z^2\right),
    \label{eq:2F1-quadratic}
\end{align}
and
\begin{align}
    {}_2F_1\left(a,a+\frac{1}{2};c;z\right)=\left(\frac{1+\sqrt{1-z}}{2}\right)^{-2 a} \,{}_2F_1\left(2a,2a-c+1;c;\frac{1-\sqrt{1-z}}{1+\sqrt{1-z}}\right).
    \label{eq:2F1-quadratic-app}
\end{align}
Eq. (\ref{eq:2F1-connection}) is valid for non-integral $a-b$, $|{\rm arg}(1-z)| < \pi$, and $|{\rm arg}(-z)| < \pi$, whereas Eq. (\ref{eq:2F1-quadratic}) is valid for $2 b \neq -1, -3, -5,\dots$ and, crucially, $\left|z\right| < 1$. Eq. (\ref{eq:2F1-quadratic-app}) is valid so long as the condition $|{\rm arg}(1-z)| < \pi$ is satisfied.

We also require some reduction identities, two for the generalized hypergeometric function ${}_3F_2$ and one for the Appell $F_1$ function. The Saalsch{\"u}tz summation formula~\cite{Slater:1966},
\begin{align}
    {}_3F_2\left(a_1,a_2,a_3;b_1,b_2;1\right)= \frac{\Gamma(b_1)\Gamma(1+a_1-b_2)\Gamma(1+a_2-b_2)\Gamma(1+a_3-b_2)}{\Gamma(1-b_2)\Gamma(b_1-a_1)\Gamma(b_1-a_2)\Gamma(b_1-a_3)},
    \label{eq:Saalschutz}
\end{align}
applies if $b_1+b_2-a_1-a_2-a_3 = 1$ and one element of $\{a_1,a_2,a_3\}$ is a negative integer. A more non-trivial summation formula involving two ${}_3F_2$ functions on the left-hand side is~\cite{WolframFunctions}
\begin{align}
    &-\frac{\Gamma(1-a_2)\Gamma(1+a_3)\Gamma(a_3-a_1)\Gamma(b_1)}{\Gamma(a_3)\Gamma(1+a_1-a_2)\Gamma(1-a_1+a_3)\Gamma(b_1-a_1)} \times \nonumber \\
    &\times {}_3F_2\left(a_1,a_1-a_3,1+a_1-b_1;1+a_1-a_2,1+a_1-a_3;1\right) + {}_3F_2\left(a_1,a_2,a_3;b_1,1+a_3;1\right) \nonumber \\
    &= \frac{\Gamma(b_1)\Gamma(1-a_2)\Gamma(1+a_3)\Gamma(a_1-a_3)}{\Gamma(a_1)\Gamma(1-a_2+a_3)\Gamma(b_1-a_3)}.
    \label{eq:Saalschutz-gen}
\end{align}
Eq. (\ref{eq:Saalschutz-gen}) is valid for $\mathfrak{Re}(1+b_1-a_1-a_2) > 0$ and may be verified using the techniques described in~\cite{Koornwinder:1998}. Finally, the Appell $F_1$ function collapses to a ${}_2F_1$ if $c = b_1 + b_2$~\cite{Schlosser:2013},
\begin{align}
    F_1\left(a;b_1,b_2;b_1+b_2;z_1,z_2\right)=(1-z_2)^{-a} \, {}_2F_1\left(a,b_1;b_1+b_2;\frac{z_1-z_2}{1-z_2}\right).
    \label{eq:F1-reduction}
\end{align}

The standard integral representation given above for the generalized hypergeometric function ${}_3F_2$ is only one of a number of Euler integrals involving ${}_2F_1$ which may be evaluated using the ${}_3F_2$ series. Many evaluations of such generalized Euler integrals are given in reference~\cite{Letessier:1988}. The main result of interest to us is
\begin{align}
    &\int _0^1 dt~ t^{\gamma -1} (1-t)^{\rho -1} (1-t z)^{-\sigma } {}_2F_1(\alpha ,\beta ;\gamma ;t) 
    = \frac{\Gamma (\gamma ) \Gamma (\rho ) \Gamma (\gamma +\rho -\alpha -\beta) }{\Gamma (\gamma +\rho -\alpha) \Gamma (\gamma +\rho -\beta)} \times \nonumber \\&~
    \qquad \qquad \times (1-z)^{-\sigma }{}_3F_2\left(\rho ,\sigma ,\gamma +\rho-\alpha -\beta; \gamma +\rho -\alpha, \gamma +\rho-\beta;\frac{z}{z-1}\right),
    \label{eq:2F1to3F2-int}
\end{align}
which is valid for $|{\rm arg}(1-z)| < \pi$, $\mathfrak{Re}(\gamma) > 0$, $\mathfrak{Re}(\rho) > 0$, and $\mathfrak{Re}(\gamma+\rho-\alpha-\beta) > 0$~\cite{Gradshteyn:2007}.


\section{All-order-in-$\ep$ results for selected purely-virtual Feynman integrals}
\label{app:virtual}
In this appendix, we collect some useful results from the Feynman integral literature. All results which follow are presented in the normalization of Eq. (\ref{eq:gen-virt-int}) and are valid to all orders in the parameter of dimensional regularization, $\ep$, for generic phase-space points in physical kinematics. In order to compare with the cut calculations performed in Section \ref{sec:stdcuts}, we must explain how to parse the ${\rm Disc}$ operation introduced in Section \ref{sec:genform}. The direct discontinuity of Feynman integral $I$ in the $s$-channel is simply
\begin{align}
{\rm Disc}_{s}(I) = I\left(s+i 0; \{v_j\}\setminus s\right) - I\left(s-i 0; \{v_j\}\setminus s\right),
\end{align}
where $\{v_j\}$ denotes the set of variables (the parameter of dimensional regularization, generalized Mandelstam variables, and, in general, internal masses) that $I$ is a function of.  


\subsection{The one-external-mass one-loop triangle}
\label{sapp:oneloop-tri}
The one-external-mass one-loop triangle with generic propagator exponents is given by~\cite{Smirnov:2004ym}
\begin{align}
    \Graph{.3}{virtonelooptriangle}    = \frac{e^{i \pi \ep} s^{2-\nu-\ep} \Gamma(2-\nu_1-\nu_3-\ep) \Gamma(2-\nu_2-\nu_3-\ep) \Gamma(\nu-2+\ep)}{\Gamma(\nu_1) \Gamma(\nu_2) \Gamma(4-\nu-2\ep)},
    \label{eq:one-loop-triangle-virt}
\end{align}
to all orders in $\ep$, where $s = (p_1+p_2)^2 > 0$ and $\nu = \sum_{i=1}^3\nu_i$.


\subsection{The massless one-loop box}
\label{sapp:oneloop-box}
The massless one-loop box is given by~\cite{vanNeerven:1985xr,Smirnov:2004ym}
\begin{align}
    \Graph{.3}{virtoneloopbox} &= - \frac{\Gamma^2(-\ep) \Gamma(\ep)}{s u \Gamma(-2\ep)}
        \bigg[ (-u)^{-\ep} {}_2 F_1\left(1, -\ep; 1-\ep; 1 + \frac{u}{s} \right) \nonumber \\&~
        + e^{i \pi \ep} s^{-\ep} {}_2 F_1\left(1, -\ep; 1-\ep; 1 + \frac{s}{u} \right)\bigg]
    \label{eq:one-loop-box-virt}
\end{align}
to all orders in $\ep$, where $s = (p_1 + p_2)^2$ and $u = (p_1 - p_4)^2$. In the physical region, $s > 0$ and $-s < u < 0$.


\subsection{The one-external-mass six-line two-loop double triangle}
The purely-virtual counterpart of the one-external-mass six-line two-loop double triangle integral studied in Section \ref{sec:stdcuts} is given by
\begin{align}
    \Graph{.3}{virttwolooptriangle} &= \frac{e^{2 i \pi \ep} s^{-2-2 \ep} \Gamma(-1-2\ep)\Gamma(1+2\ep)\Gamma^2(-\ep)}{\Gamma(1-2\ep)\Gamma(-3\ep)}\Big(\Gamma(-\ep) - \Gamma(1+\ep)\Gamma(-2\ep)\Big)
    \label{eq:two-loop-triangle-virt}
\end{align}
to all orders in $\ep$, where $s = (p_1+p_2)^2 > 0$. To our knowledge, Eq. (\ref{eq:two-loop-triangle-virt}) was first derived by van Neerven~\cite{vanNeerven:1985xr}. His idea was to first calculate all $s$-channel cuts and then deduce the associated purely-virtual result using unitarity. Consequently, to obtain (\ref{eq:two-loop-triangle-virt}) without referring to cuts, it was necessary for us to evaluate the integral ourselves using Feynman parameters. This exercise is elementary and may be carried out using the loop-by-loop integration strategy suggested in~\cite{Gonsalves:1983nq}.


\section{Baikov-Lee for the purely-virtual one-external-mass one-loop bubble}
\label{app:Baikov}
In this appendix, we complete the calculation of the Euclidean purely-virtual one-loop bubble integral with no internal masses which was initiated in Section \ref{sec:genform}. Our point of departure will be Eq. (\ref{eq:one-loop-bubble-explicit}),
\begin{align}
\Graph{.3}{virtoneloopbubble} = \int_0^{\infty} {\rm d}(q_1^2)\int_{-\sqrt{p^2 q_1^2}}^{\sqrt{p^2 q_1^2}}{\rm d}(q_1 \cdot q_2)
\frac{\pi^{\frac{3}{2}-\ep}\left(p^2 q_1^2-(q_1\cdot q_2)^2\right)^{\frac{1}{2}-\ep}}{\Gamma\left(\frac{3}{2}-\ep\right)\left(p^2\right)^{1-\ep} q_1^2\left(q_1^2-2 q_1 \cdot q_2 + p^2\right)}.
\label{eq:one-loop-bubble-explicit-recap}
\end{align}
The first step is to map the domain of the first integration variable, $q_1 \cdot q_2$, onto the unit interval. This can be achieved straightforwardly by making the change of variables $q_1 \cdot q_2 = 2 \sqrt{p^2 q_1^2} z - \sqrt{p^2 q_1^2}$.\footnote{A variable change of this form often allows one to recognize the definite integrals which arise from all-orders-in-$\ep$ Feynman integral calculations as Euler integrals of hypergeometric type (see {\it e.g.}~\cite{Gehrmann-DeRidder:2003pne}).} We arrive at
\begin{align}
\Graph{.3}{virtoneloopbubble} = \int_0^{\infty} {\rm d}(q_1^2)\int_0^1{\rm d}z
\frac{2^{2-2\ep}\pi^{\frac{3}{2}-\ep}\left(q_1^2\right)^{-\ep}(z(1-z))^{\frac{1}{2}-\ep}}{\Gamma\left(\frac{3}{2}-\ep\right)
\left(q_1^2+2(1-2 z) \sqrt{p^2 q_1^2}+p^2\right)}.
\label{eq:one-loop-bubble-varchange}
\end{align}
By comparing Eq. (\ref{eq:one-loop-bubble-varchange}) to the form of Eq. (\ref{eq:2F1-intdef}), it is now obvious that the $z$ integral may be evaluated in terms of the ${}_2F_1$ series. 

The one-fold integral that remains,
\begin{align}
\Graph{.3}{virtoneloopbubble} = \int_0^{\infty} {\rm d}(q_1^2) \frac{\pi^{2-\ep}\left(q_1^2\right)^{-\epsilon}}{\Gamma(2-\ep)(q_1^2+p^2)} 
{}_2F_1\left(1,\frac{1}{2};2-\ep;\frac{4 q_1^2 p^2}{(q_1^2+p^2)^2}\right),
\end{align}
is most naturally evaluated by splitting the integral at the point $q_1^2 = p^2$ and then mapping both the integral from $0$ to $p^2$ and the integral from $p^2$ to $\infty$ onto the unit interval. This will allow for the simultaneous application of quadratic transformation (\ref{eq:2F1-quadratic-app}) to both integrals. Making the change of variables $q_1^2 = p^2 x$ in the first integral and the change of variables $q_1^2 = p^2/x$ in the second integral, we find
\begin{align}
\Graph{.3}{virtoneloopbubble} &= \int_0^1 {\rm d}x \frac{\pi^{2-\ep}\left(p^2\right)^{-\ep}}{\Gamma(2-\ep)(1+x)}\Big(x^{-\ep}+x^{-1+\ep}\Big)
{}_2F_1\left(1,\frac{1}{2};2-\ep;\frac{4 x}{(1+x)^2}\right)\nonumber \\
&= \int_0^1 {\rm d}x \frac{\pi^{2-\ep}\left(p^2\right)^{-\ep}}{\Gamma(2-\ep)}\Big(x^{-\ep}+x^{-1+\ep}\Big)
{}_2F_1\left(1,\ep;2-\ep;x\right).
\end{align}

At this stage, we can straightforwardly evaluate both integrals using Eq. (\ref{eq:3F2-intdef}). The result obtained in this manner, 
\begin{align}
\Graph{.3}{virtoneloopbubble} &= \frac{\pi^{2-\ep}\left(p^2\right)^{-\ep}}{\Gamma(2-\ep)}
\bigg(\frac{1}{\ep}~ {}_3F_2\left(1,\ep,\ep;2-\ep,1+\ep;1\right) \nonumber \\
&\qquad+\frac{1}{1-\ep}~ {}_3F_2\left(1,\ep,1-\ep;2-\ep,2-\ep;1\right)\bigg),
\label{eq:one-loop-bubble-ugly-ans}
\end{align}
is correct but far more complicated than it needs to be. In this case, we can simplify the result by applying summation formula (\ref{eq:Saalschutz-gen}) to eliminate the second ${}_3F_2$ series in Eq. (\ref{eq:one-loop-bubble-ugly-ans}) above:
\begin{align}
\Graph{.3}{virtoneloopbubble} = \frac{\pi^{2-\ep}\left(p^2\right)^{-\ep}\Gamma^2(1-\ep)\Gamma(\ep)}{\Gamma(2-2\ep)}.
\end{align}
Needless to say, the above result agrees with what one obtains (far more easily) using the Feynman representation.\footnote{One can see that this is the case by consulting a standard text such as Smirnov~\cite{Smirnov:2004ym}.}


\bibliographystyle{JHEP}
\bibliography{refs.bib}{}
\end{document}